\pdfoutput=1

\documentclass[11pt,twoside,a4paper,cmspaper,final,collab]{cms-tdr}
\def\svnVersion{167113}\def\cmsCernNo{2012-123}\def\cmsCernDate{\today}\def\cmsMessage{Submitted to the Journal of High Energy Physics}

\begin{document}\cmsNoteHeader{HIG-11-019}

\hyphenation{had-ron-i-za-tion}
\hyphenation{cal-or-i-me-ter}
\hyphenation{de-vices}

\RCS$Revision: 134545 $
\RCS$HeadURL: svn+ssh://svn.cern.ch/reps/tdr2/papers/HIG-11-019/trunk/HIG-11-019.tex $
\RCS$Id: HIG-11-019.tex 134545 2012-07-03 13:38:20Z anikiten $
\newcommand{\mt}{\ensuremath{m_\mathrm{T}}}
\cmsNoteHeader{HIG-11-019} 
\title{Search for a light charged Higgs boson in top quark decays in $\Pp\Pp$ collisions at $\sqrt{s} = 7\TeV$}

\date{\today}

\abstract{
   Results are presented on a search for a light charged Higgs boson
   that can be produced in the decay of the top quark $\cPqt \rightarrow \PH^{+}\cPqb$
   and which, in turn, decays into $\Pgt^{+} \Pgngt$. The analysed data correspond to
   an integrated luminosity of about 2\fbinv recorded in proton-proton collisions
   at $\sqrt{s} = 7\TeV$ by the CMS experiment at the LHC. The search is sensitive to the decays of the top quark
   pairs $\ttbar \rightarrow \PHpm W^{\mp} \cPqb \cPaqb$ and $\ttbar \rightarrow \PHpm \PH^{\mp} \cPqb \cPaqb$.
   Various final states have been studied separately, all requiring presence of a $\Pgt$ lepton from
   $\PH^{+}$ decays, missing transverse energy, and multiple jets. Upper limits on the branching fraction
   $\mathcal{B}({\cPqt} \rightarrow \PH^{+} \cPqb)$ in the range of 2--4\% are established
   for charged Higgs boson masses  between 80 and 160\GeV, under the assumption that
   $\mathcal{B}(\PH^{+} \rightarrow \Pgt^{+} \Pgngt) = 1$.}
\hypersetup{%
pdfauthor={CMS Collaboration},%
pdftitle={Search for a light charged Higgs boson in top quark decays in pp collisions at sqrt(s) = 7 TeV},%
pdfsubject={CMS},%
pdfkeywords={CMS, physics, Higgs, tau, MSSM}}

\maketitle 

\section{Introduction \label{sec:introduction}}
The minimal supersymmetric extension of the standard model (MSSM) requires the introduction of two Higgs doublets
in order that the superpotential can contain appropriate terms for giving masses to both up and down type quarks~\cite{Fayet:1974pd,Fayet:1976et,Fayet:1977yc,Dimopoulos:1981zb,Sakai:1981gr,Inoue:1982ej,Inoue:1982pi,Inoue:1983pp}.
This leads to the prediction of five elementary Higgs particles: two CP-even ($\mathrm{h}$,$\PH$), one CP-odd ($\mathrm{A}$), and two charged ($\PHpm$)
states ~\cite{Gunion:1989we,Djouadi:2005gj}. The lower limit on the charged Higgs boson mass is $78.6\GeV$, as determined by LEP
experiments~\cite{Achard:2003gt,Heister:2002ev,Abdallah:2003wd,Abbiendi:2008aa}.
If the mass of the charged Higgs boson is smaller than the difference between the masses of
the top and the bottom quarks, \ie $m_{\PH^{+}}<m_{\cPqt}-m_{\cPqb}$, the top quark can decay via $\cPqt\rightarrow \PH^{+} \cPqb$
(charge conjugate processes are always implied throughout this paper).
For values of $\tan \beta > 5$, the charged Higgs boson preferentially decays to a $\Pgt$ lepton and a neutrino,
$\PH^{+}\rightarrow \Pgt^{+} \Pgngt$, where $\tan \beta$ is defined as the ratio of the vacuum expectation values of the two Higgs
boson doublets. In deriving the experimental limits we assume that the branching fraction
$\mathcal{B}(\PH^{+} \rightarrow \Pgt^{+}\Pgngt)$ is equal to 1.

The presence of the $\cPqt \rightarrow \PH^{+} \cPqb$, $\PH^{+}$$\rightarrow \Pgt^{+} \Pgngt$ decay modes alters
the $\Pgt$ lepton yield in the decay products of $\ttbar$ pairs compared to the standard model (SM). The upper limit
on the branching fraction, $\mathcal{B}(\cPqt\rightarrow \PH^{+}\cPqb) < 0.2$, has been set by the CDF~\cite{cdfchiggs2006} and D0~\cite{d0chiggs}
experiments at the Tevatron for $m_{\PH^{+}}$ between 80 and 155\GeV, assuming $\mathcal{B}(\PH^{+}$$\rightarrow \Pgt^{+} \Pgngt) = 1$.
More recently, ATLAS experiment at the LHC has set the upper limit on the $\mathcal{B}(\cPqt\rightarrow \PH^{+}\cPqb)$ between 5\% and 1\% for
charged Higgs boson masses in the range 90--160\GeV~\cite{atlasH+}.

The dominant process of production of top quarks at the Large Hadron Collider (LHC) is $\Pp\Pp\rightarrow \ttbar$+X via gluon gluon fusion.
The search for a charged Higgs boson is sensitive to the decays of the top quark pairs
$\ttbar \rightarrow \PHpm W^{\mp} \cPqb \cPaqb$ and $\ttbar \rightarrow \PHpm \PH^{\mp} \cPqb \cPaqb$,
where each charged Higgs boson decays into a $\Pgt$ lepton and a neutrino. Throughout this paper, these two decay modes
are referred to as $\PW\PH$ and $\PH\PH$, respectively.

\begin{figure*}[htb]
\begin{center}
\includegraphics[width=0.3\textwidth]{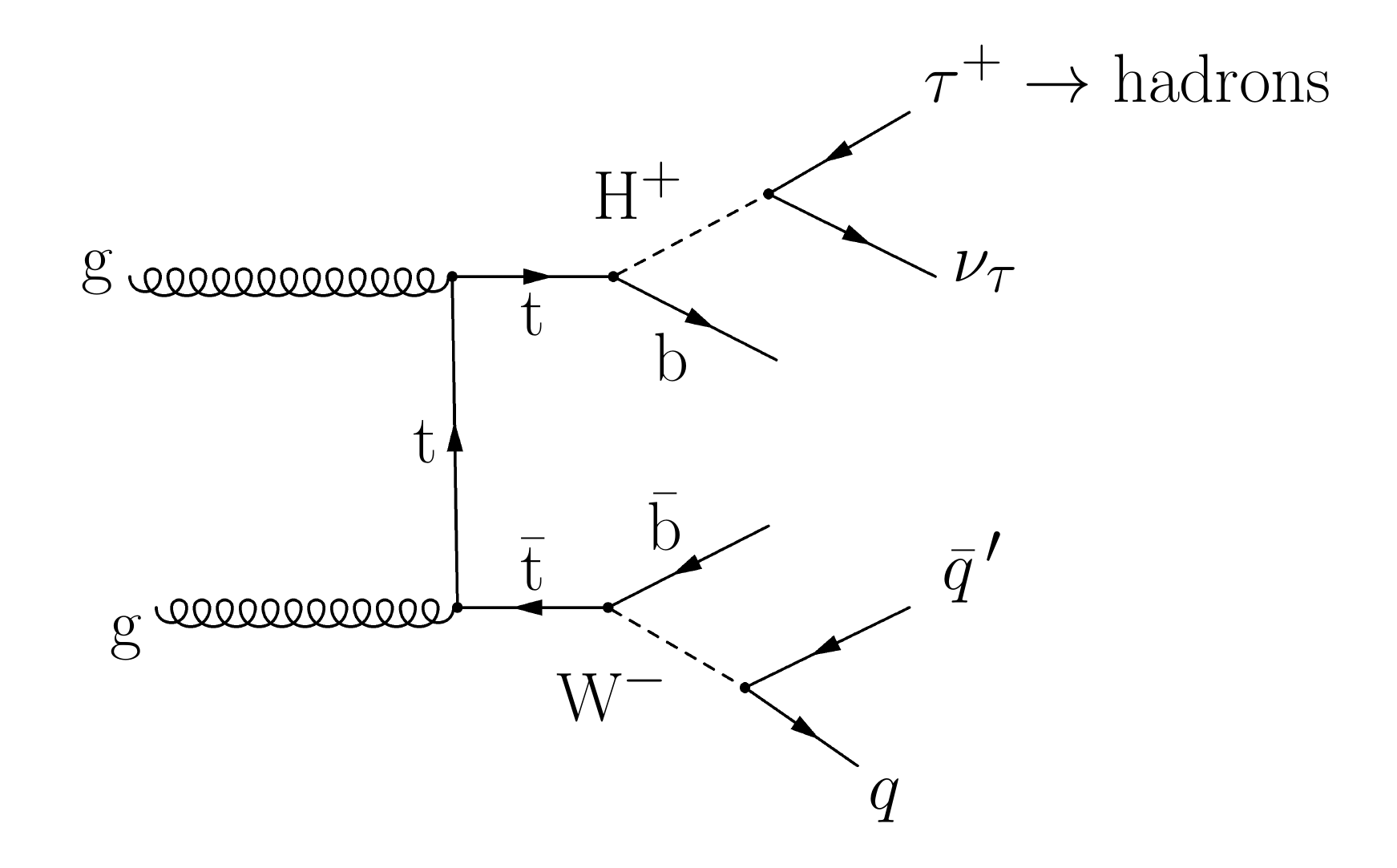}
\includegraphics[width=0.3\textwidth]{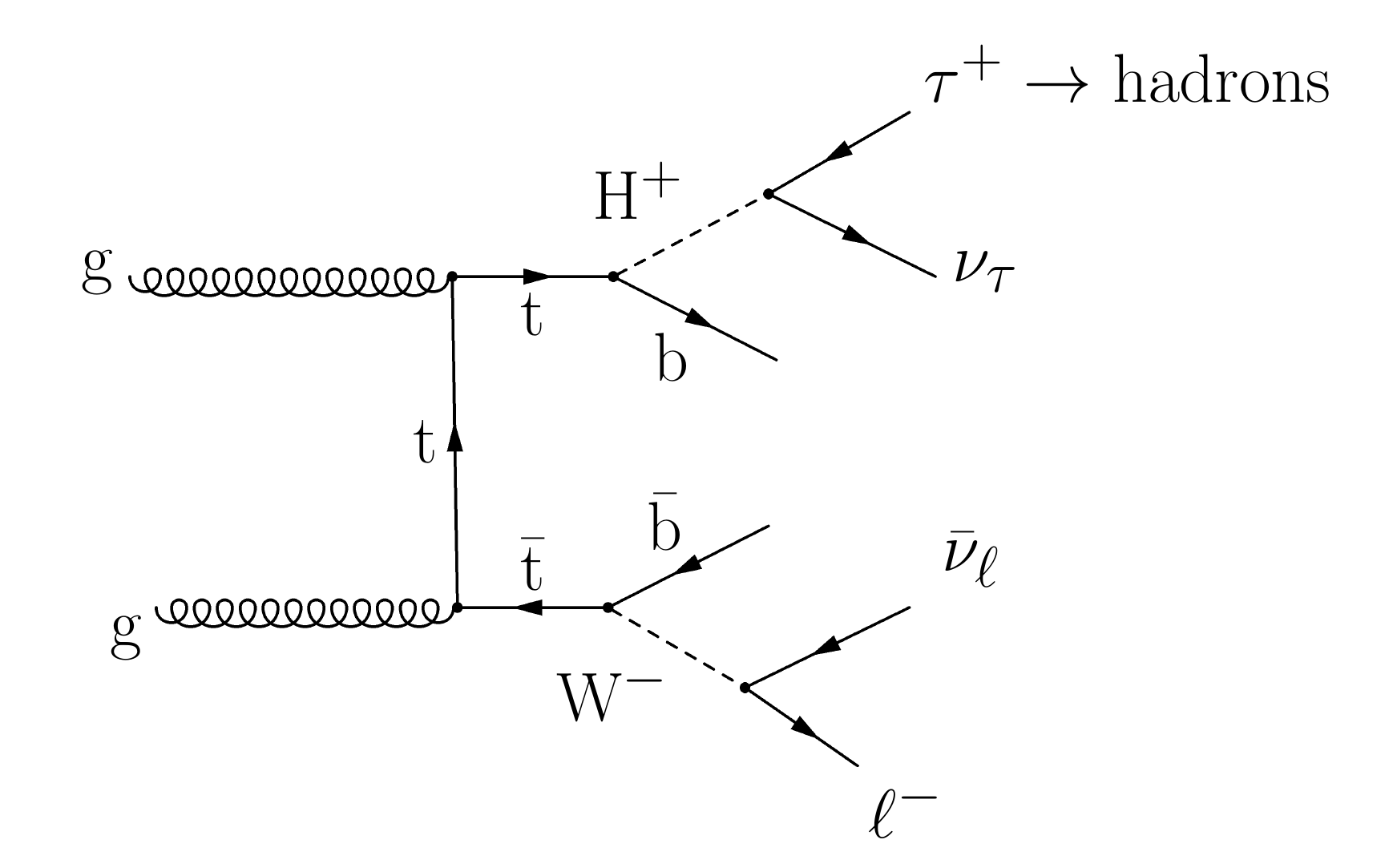}
\includegraphics[width=0.3\textwidth]{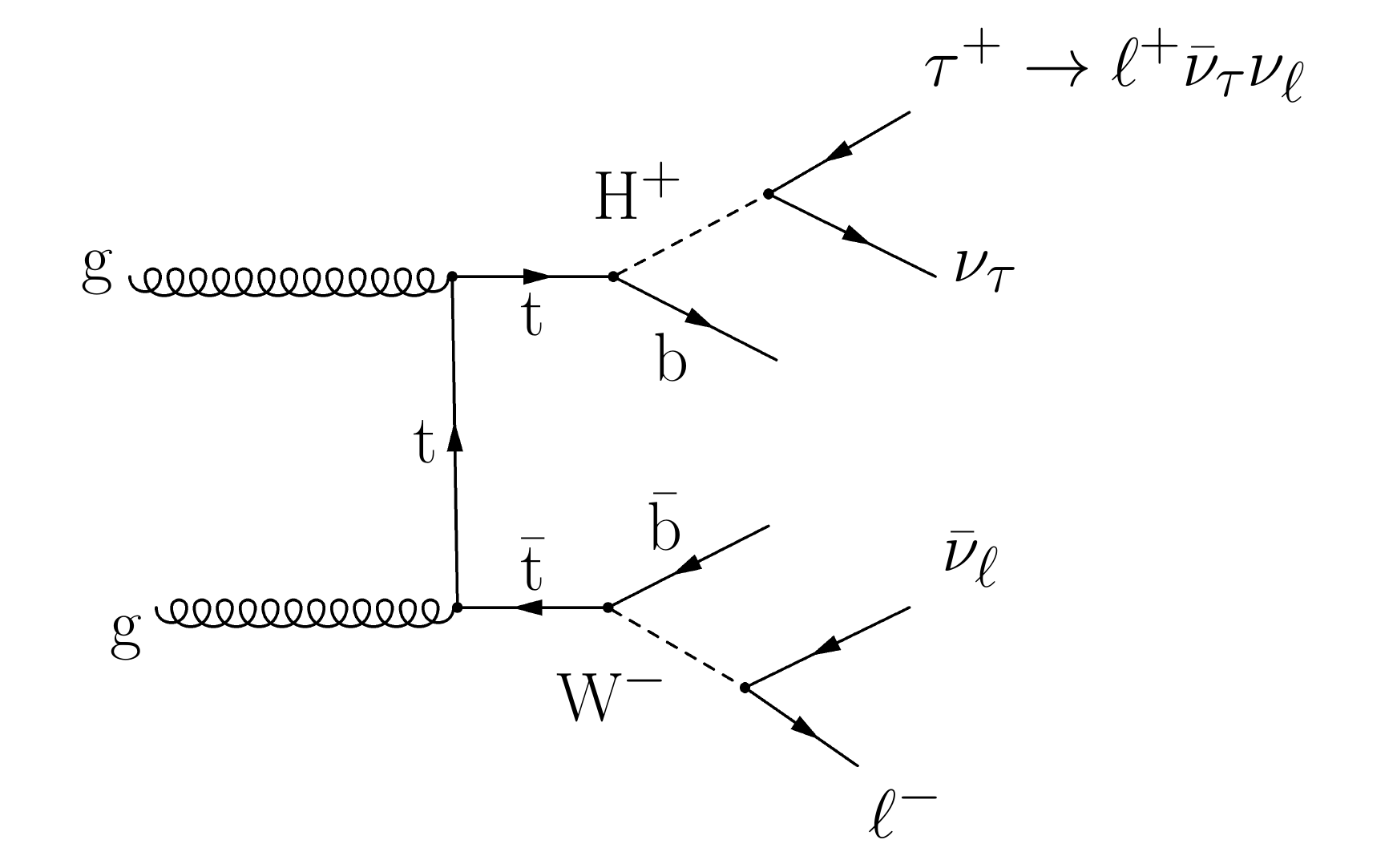}
\caption{Representative diagrams for the $\Pgt_\mathrm{h}$+jets (left), $\Pe(\Pgm)\Pgt_\mathrm{h}$ (middle), and $\Pe\Pgm$ (right) final states.}
\label{fig:introduction}
\end{center}
\end{figure*}

Three different final states are studied, all requiring missing transverse energy and multiple jets.
The $\Pgt$ lepton decaying into hadrons and a neutrino is labeled $\Pgt_\mathrm{h}$. The first final state involves
the production of $\Pgt_\mathrm{h}$ and jets (labeled $\Pgt_\mathrm{h}$+jets), the second one is where $\Pgt_\mathrm{h}$ is produced in association
with an electron or a muon (labeled $\Pe\Pgt_\mathrm{h}$ or $\Pgm\Pgt_\mathrm{h}$), and the third one is where an electron and a muon are produced
(labeled $\Pe\Pgm$). Figure~\ref{fig:introduction} shows representative diagrams for the $\Pgt_\mathrm{h}$+jets (left plot), $\Pe(\Pgm)\Pgt_\mathrm{h}$
(middle plot), and $\Pe\Pgm$ (right plot) final states. We use a data sample recorded by the Compact Muon Solenoid (CMS) experiment until the
end of August 2011 with an average number of interactions per crossing (pileup) of 5--6.
The analyses correspond to an integrated luminosity ranging from 1.99 to 2.27 fb$^{-1}$ depending on the final state.

\section{CMS detector, reconstruction, and simulation \label{sec:cmsdetrecsim}}
A detailed description can be found in Ref.~\cite{:2008zzk}.
The central feature of the CMS apparatus is a superconducting solenoid of 6\unit{m} internal diameter providing a field of
3.8\unit{T}. Within the field volume are a silicon pixel and strip tracker, a crystal electromagnetic calorimeter (ECAL), and a brass/scintillator
hadron calorimeter (HCAL). Muons are measured in gas-ionization detectors embedded in the steel return yoke of the magnet. Extensive forward calorimetry
complements the coverage provided by the barrel and endcap detectors.

CMS uses a right-handed coordinate system, with the origin at the nominal interaction point, the $x$ axis pointing to the centre of the LHC,
the $y$ axis pointing up (perpendicular to the LHC plane), and the $z$ axis along the anticlockwise-beam direction. The polar angle $\theta$
is measured from the positive $z$ axis and the azimuthal angle $\phi$ is measured in the $x$-$y$ plane. The preudorapidity $\eta$ is defined as
$-\ln[\tan(\theta/2)]$.

The first level (L1) of the CMS trigger system, composed of custom hardware processors, uses information from the calorimeters and muon detectors
to select the most interesting events in a fixed time interval of less than 4\mus. The High Level Trigger (HLT) processor farm further decreases
the event rate from around 100\unit{kHz} to around 300\unit{Hz}, before data storage.

Muons are reconstructed~\cite{muon} by performing a simultaneous global track fit to hits in the silicon tracker and the muon system.
Electrons are reconstructed~\cite{electron} from clusters of energy deposits in the electromagnetic calorimeter that are matched
to hits in the silicon tracker. Jets, $\Pgt_\mathrm{h}$, and missing transverse energy ($\ETmiss$) are reconstructed
using particles measured with the particle-flow algorithm~\cite{particleflow}. The particle-flow algorithm reconstructs particles in each event,
using the information from the tracker, the ECAL and HCAL calorimeters, and the muon system. Jets are reconstructed with the anti-$k_T$ jet
algorithm~\cite{Cacciari:2008gp} with a distance parameter of $R=0.5$. The value of $\ETmiss$ is defined as the magnitude of the vector sum
of the transverse momenta of all reconstructed objects in the volume of the detector (leptons, photons, and hadrons).

The b-tagging algorithm used in this analysis exploits as the discriminating variable the significance of the impact parameter
of the track with the second highest significance~\cite{btag-11-001}. The significance is defined as the ratio of the measured
value of the impact parameter to the measurement uncertainty. The hadron-plus-strips (HPS) $\Pgt$ identification
algorithm~\cite{1748-0221-7-01-P01001} is used to reconstruct $\Pgt$ leptons decaying hadronically.  The HPS algorithm
considers candidates with one or three charged pions and up to two neutral pions. The $\Pgt_\mathrm{h}$ candidate
isolation is based on a cone of $\Delta R=\sqrt{\Delta \phi ^{2}+ \Delta \eta ^{2}} = 0.5$ around the reconstructed
$\Pgt_\mathrm{h}$-momentum direction. It is required that no charged hadrons with $\pt > \pt^\text{cut}$ and no photons with
$\ET > \ET^\mathrm{cut}$ be present within the isolation cone, other than the $\Pgt_{h}$ constituents. The typical values of
$\pt^\mathrm{cut}$ and $\ET^\mathrm{cut}$ are ${\simeq} 1\GeV$.

Backgrounds $\ttbar$, $\PW$+jets, $\cPZ$+jets are generated with \MADGRAPH5~\cite{Maltoni:2002qb,Alwall:2011uj}
interfaced with \PYTHIA6.4.25~\cite{Sjostrand:2006za}. The diboson production processes $\PW\PW$, $\PW\cPZ$, and $\cPZ\cPZ$ are generated
by \PYTHIA. Single-top-quark production is generated with \POWHEG~\cite{Frixione:2007vw}. The signal processes,
$\ttbar \rightarrow \PHpm \cPqb\PH^{\mp}\cPaqb$ and $\ttbar \rightarrow \PW^{\pm}\cPqb\PH^{\mp}\cPaqb$, are generated by \PYTHIA.
The \TAUOLA~\cite{Was:2000st} package is used to simulate $\Pgt$ decays in all cases.

Generated events are processed through the full detector simulation based on \GEANTfour~\cite{Agostinelli:2002hh,Allison:2006ve}, followed
by a detailed trigger emulation and the CMS event reconstruction. Several minimum-bias events are superimposed upon the hard interactions
to simulate pileup. The simulated events are weighted according to the measured distribution of the number of interaction vertices. The \PYTHIA
parameters for the underlying event were set according to the ``Z2'' tune, an update of the ``Z1'' tune described in Ref.~\cite{Field:2010bc}.

The number of produced $\ttbar$ events is estimated from the SM prediction of the $\ttbar$ production cross section,
$165^{+4}_{-9}$(scale)$^{+7}_{-7}$(PDF)\unit{pb}~\cite{Langenfeld:2009tc,Kidonakis:2010dk,Cacciari:2008zb,Dittmaier:2012vm}.
The theoretical prediction agrees with the cross section measured at the LHC~\cite{PhysRevD.84.092004,ATLAS:2012aa}.
\section{Analysis of the \texorpdfstring{$\Pgt_\mathrm{h}$+jets}{tau(h)+jets} final state \label{sec:taujets}}
In the $\Pgt_\mathrm{h}$+jets analysis, events are selected by a trigger that requires the presence of a $\Pgt_\mathrm{h}$ with
transverse momentum $\pt > 35\GeV$ and a large calorimetric $\ETmiss$ ($> 60\GeV$). The $\Pgt_\mathrm{h}$ trigger
selection includes the requirement on the leading-$\pt$ track, $\pt> 20\GeV$. The amount of data
analyzed for this channel corresponds to an integrated luminosity of $2.27 \pm 0.05\fbinv$.

In this analysis, selected event are required to have one $\Pgt_\mathrm{h}$ with
$\pt^{\Pgt_\mathrm{h}} > 40\GeV$ within $| \eta | < 2.1$, and at least three other
jets with $\pt > 30\GeV$ within $| \eta | < 2.4$ with at least one jet identified as originating from the hadronization of a b quark.

In order to suppress the multijet background we use selection criteria
on the missing transverse energy, $\ETmiss > 50\GeV$, and on the angle between the $\ETmiss$ vector and
the transverse momentum of the $\Pgt_\mathrm{h}$, $\Delta \phi(\pt^{\Pgt_\mathrm{h}},\ETmiss) < 160^{\circ}$.
This analysis selects $\Pgt_{h}$ candidates with one charged hadron. The charged hadron is required to have $\pt^\text{trk} > 20\GeV$.
In order to use non-overlapping data samples in the $\Pgt_\mathrm{h}$+jets analysis and the other analyses,
events containing either an electron or a muon with $\pt^{\ell} > 15\GeV$ are rejected.
The background events with $\PW\rightarrow\Pgt\Pgngt$ decays are suppressed by a requirement on the variable
$R_{\Pgt} = p^\text{trk}/p^{\Pgt_\mathrm{h}}$, with $R_{\Pgt} > 0.7$, which takes into account the different polarization of
$\Pgt$ leptons originating from $\PH$ or $\PW$ decays~\cite{Roy:1999xw}.
Although the requirements on the transverse momenta of $\Pgt_\mathrm{h}$ and the charged particle introduce a bias
of $R_{\Pgt}$ requirement, it provides a background rejection factor of about two.

In the $\Pgt_\mathrm{h}$+jets analysis the dominant reducible background arises from multijet events with large $\ETmiss$ and
jets that mimic hadronic $\Pgt$ decays or are misidentified as \cPqb-quark jets.

The other background processes comprise electroweak (EWK) ones - $\PW$+jets, $\cPZ$+jets, diboson ($\PW\PW$, $\cPZ\cPZ$, $\PW\cPZ$) as well as
SM $\ttbar$ and $\cPqt\PW$ production. The $\PW$+jets and $\ttbar$ production processes dominate. These backgrounds can be divided
in two parts:  the first one labeled ``EWK+$\ttbar$~$\Pgt$" consists of events where at least one $\Pgt$ lepton in the final state is present with
$\pt^{\Pgt} > 40\GeV$, within $|\eta^{\Pgt}| < 2.1$, and the second one labeled ``EWK+$\ttbar$~no-$\Pgt$" consisting of events
with no $\Pgt$ leptons in the final state or with no $\Pgt$ leptons satisfying the above-mentioned criteria. The ``EWK+$\ttbar$~no-$\Pgt$"
background events with no $\Pgt$ leptons in the final state can pass the selection due to misidentification of a jet, an electron or a
muon as a $\Pgt_\mathrm{h}$.

The transverse mass, $\mt$ , can be reconstructed from the $\Pgt_\mathrm{h}$ and $\ETmiss$ vectors,
providing additional discrimination between $\PW$ and $\PH$ decays.
The shape and normalisation of the $\mt$  distributions of the multijet and ``EWK+$\ttbar$~$\Pgt$" backgrounds are obtained from data.
The $\mt$  distribution of the multijet background is measured using the events which pass the signal selection described above,
except for no requirements on $\Pgt$ isolation and on an identified b quark jet. A small contamination from EWK+$\ttbar$ processes,
evaluated using simulation, has been subtracted. The $\mt$  distributions are measured in bins of $\pt$ of
the $\Pgt$ candidate (a $\Pgt_{h}$ with no isolation criteria applied).
The final $\mt$  distribution of the multijet background after full event selection is obtained by summing the $\mt$
distributions for each $\pt^{\Pgt}$ bin weighted with the efficiency that the $\Pgt$ candidate passes the $\Pgt$ isolation
criteria and the $R_{\Pgt}$ selection. The efficiency is measured from data using events selected for the measurement of the
$\mt$  distributions, but without applying the requirements on $\ETmiss$ and $\Delta \phi(\pt^{\Pgt_\mathrm{h}}, \ET^\text{miss})$.
The expected number of multijet events in a given bin $i$ of the $\mt$  distribution is calculated as:

\begin{equation} \label{eq:qcd1}
N_{i}^{\text{multijet}} = N^{\text{multijet}} \sum_{j} p_{i,j}^{\text{multijet}} \,\varepsilon _{j}^{\Pgt+R_{\Pgt}},
\end{equation}
where the index $j$ runs over the bins of $\pt^{\Pgt_\mathrm{h}}$; $\varepsilon _{j}^{\Pgt+R_{\Pgt}}$ is the efficiency of the
$\Pgt$ isolation and the $R_{\Pgt}$ selection in a given bin $j$, $p_{i,j}^\text{multijet}$ is the $\mt$  probability density
function obtained from the shapes of the $\mt$  distributions, and $N^\text{multijet}$ is the total number of the multijet background events.

The expected number of events and the $\mt$  distribution of the ``EWK+$\ttbar$~$\Pgt$" background are obtained using
a control data sample defined with the same jet selection criteria of the $\Pgt_\mathrm{h}$+jets sample, but requiring a muon instead
of a $\Pgt_\mathrm{h}$. The reconstructed muons are then replaced by embedding in the events the reconstructed particles from simulated
$\Pgt$ lepton decays. The embedding method underestimates a small contribution from the Drell-Yan $\Pgt \Pgt$ and
$\PW\PW$$\rightarrow \Pgt\Pgt + \ETmiss$
processes, since a veto on the presence of a second lepton ($e$ or $\mu$) is used in the selection of the control sample.
The residues of these backgrounds not counted with the embedding method have been estimated from the simulation.
The ``EWK+$\ttbar$~no-$\Pgt$" background has been estimated from the simulation.

\begin{figure}[htb]
\begin{center}
\includegraphics[width=0.60\textwidth]{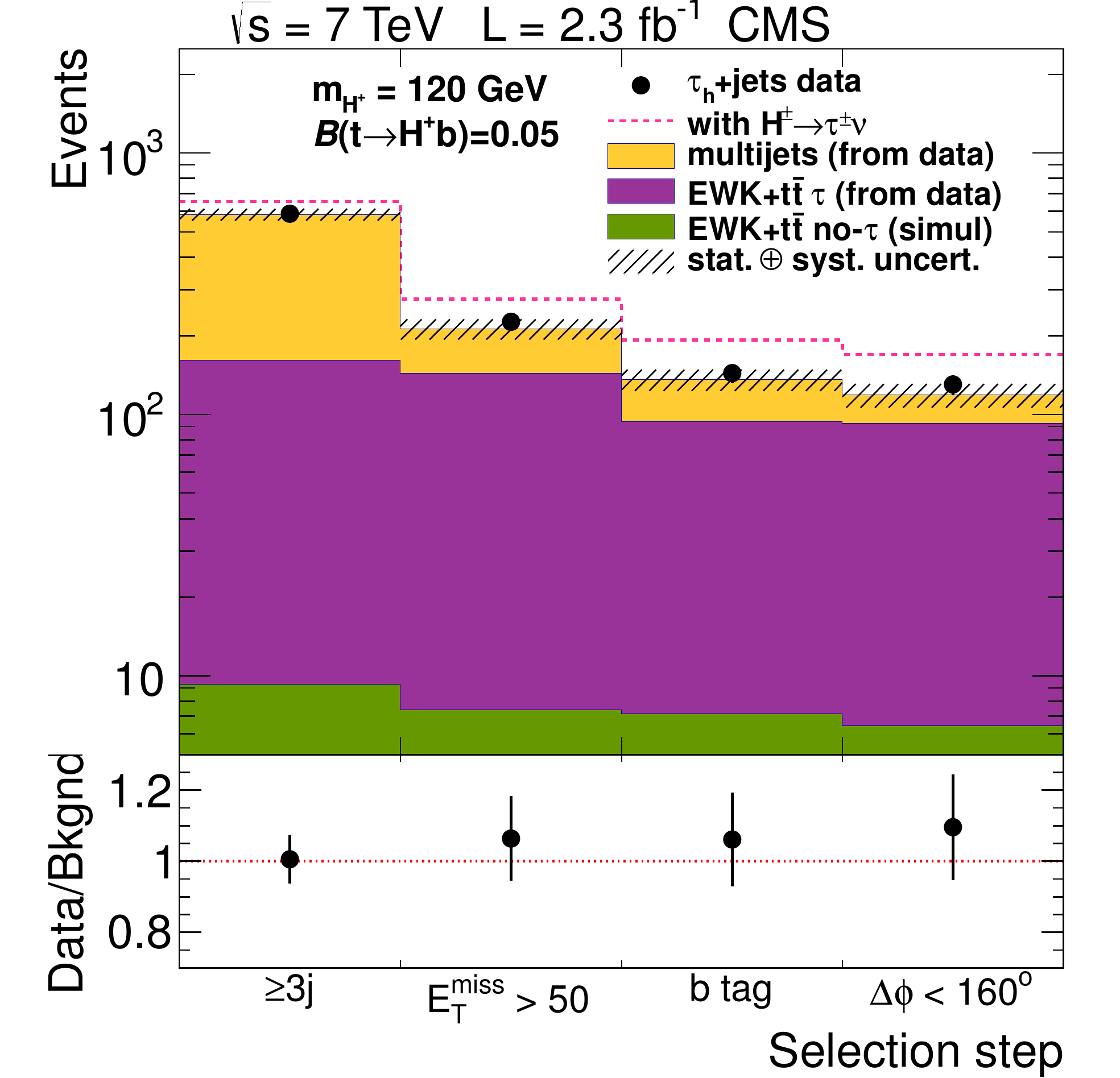}
\caption{The event yield after each selection step for the $\Pgt_\mathrm{h}$+jets analysis. The expected event yield in the presence of the
         $\cPqt\rightarrow\PH^{+}\cPqb$, $\PH^{+}\rightarrow\Pgt^{+}\Pgngt$ decays is shown as the dashed line for
         $m_{\PH^{+}} = 120\GeV$ and under that assumption that $\mathcal{B}(\cPqt\rightarrow\PH^{+}\cPqb) = 0.05$.
         The multijet and the ``EWK+$\ttbar$ $\Pgt$" backgrounds are measured from the data.
         The ``EWK+$\ttbar$ no-$\Pgt$" background is shown as estimated from simulation.
         The bottom panel shows the ratio of data over background along with the total uncertainties.
         Statistical and systematic uncertainties are added in quadrature.}
\label{fig:fullyhadronic_selections_fig1ab}
\end{center}
\end{figure}

Figure~\ref{fig:fullyhadronic_selections_fig1ab} shows the event yield after each selection step starting from
the requirement that at least three high-$\pt$ jets are present. The expected event yield in the presence of the
$\cPqt\rightarrow\PH^{+}\cPqb$, $\PH^{+}$$\rightarrow\Pgt^{+}\Pgngt$ decays is shown as the dashed line for $m_{\PH^{+}} = 120\GeV$ and
assuming $\mathcal{B}(\cPqt \rightarrow \PH^{+}\cPqb) = 0.05$. The multijet background and the ``EWK+$\rm t\bar{\cPqt}$ $\Pgt$" background are shown as
measured from the data. The ``EWK+$\ttbar$ no-$\Pgt$" background is shown as estimated from the simulation.

The observed number of events after full event selection is listed in Table~\ref{tab:fullyhadronic_selections_tab1},
along with the expected number of events from the various backgrounds, and from the Higgs boson signal processes $\PW\PH$ and $\PH\PH$
at $m_{\PH^{+}} = 120\GeV$. The number of $\PW\PH$ and $\PH\PH$ events is obtained under the assumption that
$\mathcal{B}(\cPqt\rightarrow\PH^{+}\cPqb) = 0.05$. The systematic uncertainties listed in Table~\ref{tab:fullyhadronic_selections_tab1} will be
discussed in Section~\ref{sec:systematics}.
\begin{table*}[htbp]
\begin{center}
\topcaption{Numbers of expected events in the $\Pgt_\mathrm{h}$+jets analysis for the backgrounds and the Higgs boson signal from
$\PH\PH$ and $\PW\PH$ processes at $m_{\PH^{+}} = 120\GeV$, and the number of observed events after the final event selection.
Unless stated differently, the expected background events are from simulation.}
\begin{tabular}{|c|c|}
\hline
\multicolumn{1}{|c|}{Source\rule{0pt}{12pt}} & \multicolumn{1}{c|}{$N_\mathrm{ev}^{\Pgt_\mathrm{h}+\text{jets}} \pm \text{stat.} \pm \text{syst.}$}    \\
\hline
\hline
$\PH\PH+\PW\PH$, $m_{\PH^{+}}=120\GeV$, $\mathcal{B}(\cPqt\rightarrow\PH^{+}\cPqb)=0.05$ &   51 $\pm$ 4~$\pm$ 8 \\
\hline
\hline
multijets (from data)                              &    26   $\pm$ 2    $\pm$ 1    \\
EWK+$\ttbar$~$\Pgt$ (from data)                   &    78   $\pm$ 3    $\pm$ 11   \\
EWK+$\ttbar$~no-$\Pgt$               &    6.0    $\pm$ 3.0    $\pm$ 1.2  \\
residual $\cPZ$$/\gamma^{\ast}\rightarrow\Pgt\Pgt$     &    7.0  $\pm$ 2.0  $\pm$ 2.1   \\
residual $\PW\PW\rightarrow\Pgt\Pgngt\Pgt\Pgngt$  &    0.35 $\pm$ 0.23 $\pm$ 0.09  \\
\hline
Total expected background          &  119  $\pm$ 5 $\pm$ 12   \\
\hline
\hline
Data                                &    130                              \\
\hline
\end{tabular}

\label{tab:fullyhadronic_selections_tab1}
\end{center}
\end{table*}

\begin{figure}[htb]
\begin{center}
\includegraphics[width=0.60\textwidth]{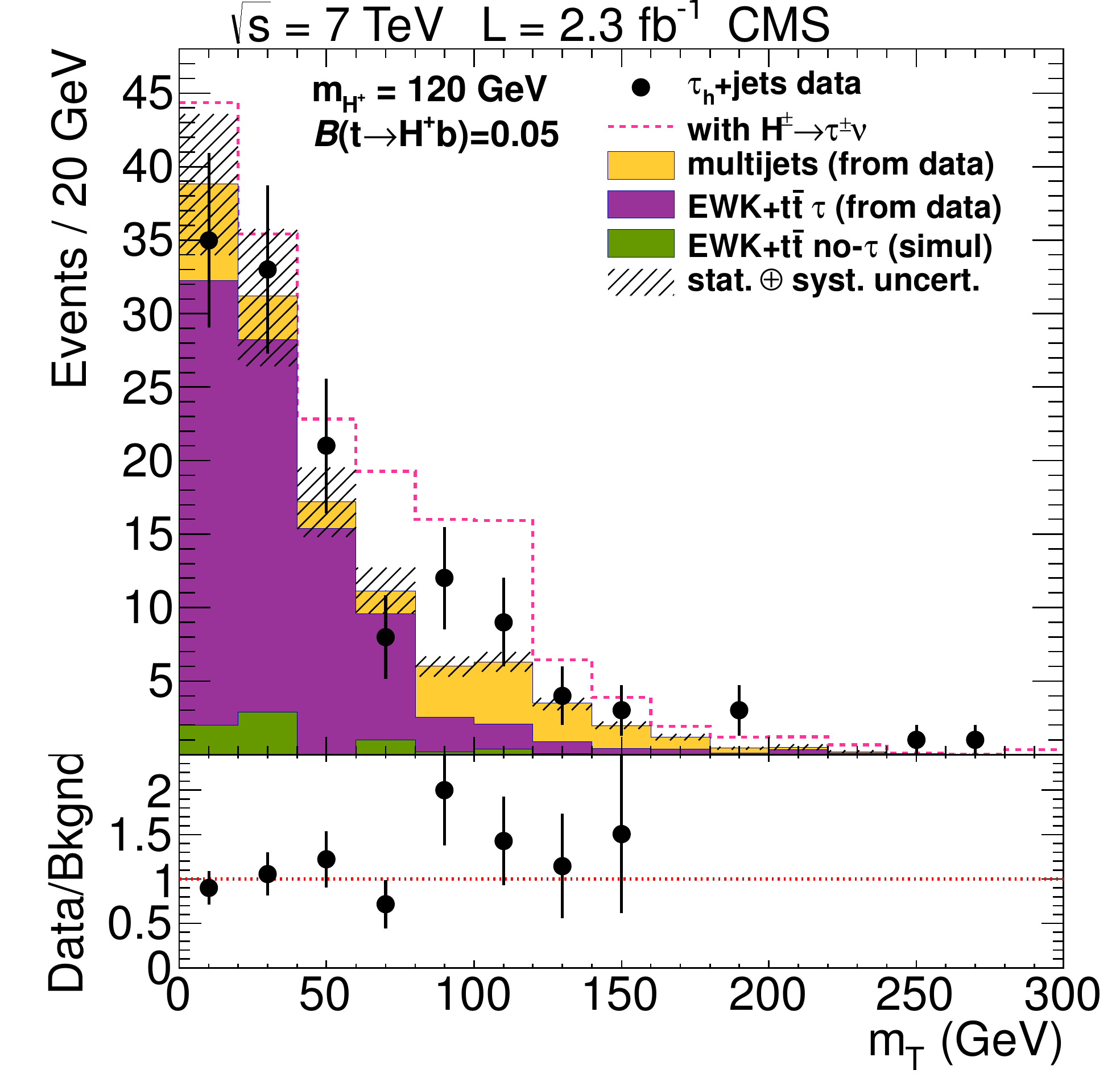}
\caption{The transverse mass of $\Pgt_\mathrm{h}$ and $\ETmiss$ after full event selection for the $\Pgt_\mathrm{h}$+jets analysis.
         The expected event yield in the presence of the
         $\cPqt\rightarrow\PH^{+}\cPqb$, $\PH^{+}\rightarrow\Pgt^{+}\Pgn$ decays is shown as the dashed line for
         $m_{\PH^{+}} = 120\GeV$ and under the assumption that $\mathcal{B}(\cPqt\rightarrow\PH^{+}\cPqb) = 0.05$.
         The bottom panel shows the ratio of data over background along with the total uncertainties.
	 The ratio is not shown for $\mt > 160\GeV$, where the expected total number of the background events
         is $2.5\pm0.3$ while 5 events are observed. Statistical and systematic uncertainties are always added
         in quadrature.}
\label{fig:fullyhadronic_selections_fig2}
\end{center}
\end{figure}

The $\mt$  distribution after all event selection criteria are applied is shown in Fig.~\ref{fig:fullyhadronic_selections_fig2}.

\section{Analysis of the \texorpdfstring{$\Pe\Pgt_\mathrm{h}$ and $\Pgm\Pgt_\mathrm{h}$}{e tau(h) and mu tau(h)} final states \label{sec:ltau}}
The event selections used are the same as in the measurement of the top quark pair production cross section
in dilepton final states containing $\Pgt$~\cite{Collaboration:2012vs}.

In the  $\Pe\Pgt_\mathrm{h}$ analysis, the events are selected by a trigger that requires the
presence of an electron, at least two jets with $\pt > 30\GeV$ and $\pt > 25\GeV$, respectively,
and a certain amount  of $H_\mathrm{T}^\text{miss}$, where $H_\mathrm{T}^\text{miss}$ is defined at the trigger level as the
magnitude of the vector sum of $\pt$ of all jets in the event. As the peak instantaneous luminosity increased the requirements
on the electron $\pt$ changed from 17 to 27 GeV and on $H_\mathrm{T}^\text{miss}$ from 15 to 20\GeV.
The amount of data analyzed for this channel corresponds to an integrated luminosity of $1.99 \pm 0.05\fbinv$

In the $\Pgm\Pgt_\mathrm{h}$ analysis, the events are selected by a single-muon trigger with the threshold
changing from 17 to 24\GeV during the data taking period. The amount of data analyzed for this channel corresponds
to an integrated luminosity of $2.22 \pm 0.05\fbinv$.

The events are selected by requiring one isolated, high-$\pt$ electron (muon) with
$\pt> 35 \,(30)\GeV$ and $|\eta| < 2.5 \,(2.1)$. The event should have one $\Pgt_\mathrm{h}$ with
$p_{T} > 20\GeV$ within $|\eta| < 2.4$, at least two jets with $\pt > 35 \,(30)\GeV$ within $|\eta| < 2.4$, with at least
one jet identified as originating from the hadronization of a $\cPqb$ quark, and $\ETmiss > 45 \,(40)\GeV$ for
the $\Pe\Pgt_\mathrm{h}$ ($\Pgm\Pgt_\mathrm{h}$) final state. The $\Pgt_\mathrm{h}$ and the electron (muon) are required to have opposite electric charges.
The isolation of each charged lepton candidate ($\Pe$ or $\Pgm$) is measured by summing the transverse momenta of the reconstructed particles
within a cone of radius $\Delta R = 0.3$ around the lepton's direction.
The contribution from the lepton itself is excluded. If the value of this sum divided by the lepton
$\pt$, labeled $I_\text{rel}$, is less than 0.1 (0.2) for electrons (muons), the lepton is considered to be isolated.
The lepton is required to be separated from any selected jet by a distance $\Delta R >$ 0.3. Events with
an additional electron (muon) with $I_\text{rel} < 0.2$ and $p_T > 15 \,(10)\GeV$ are rejected.

The backgrounds in the $\Pe\Pgt_\mathrm{h}$ and $\Pgm\Pgt_\mathrm{h}$ final-state analyses arise from two sources, the first with misidentified
$\Pgt_\mathrm{h}$, which is estimated from data, and the second with genuine $\Pgt_\mathrm{h}$, which is estimated from
simulation. The misidentified $\Pgt_\mathrm{h}$ background comes from events with one lepton ($\Pe$ or $\Pgm$), $\ETmiss$, and
three or more jets with at least one identified b quark jet (labelled ``$\ell+\geq3$~jets" events), where one jet is misidentified as a
$\Pgt_\mathrm{h}$. The dominant contribution to this background comes from $\PW$+jets, and
from $\ttbar\rightarrow\PW^+\cPqb\PW^-\cPaqb\rightarrow$~$\ell\cPgn\cPqb ~q q' \cPaqb$ ($\ell = \Pe,\Pgm$) events.
The misidentified $\Pgt_\mathrm{h}$ background
is estimated by applying the probability that a jet mimics a $\Pgt_\mathrm{h}$ to every jet in ``$\ell+\geq3$~jets" events. The probability that a
jet is misidentified as a $\Pgt_\mathrm{h}$ is measured from data as a function of jet $\pt$ and $\eta$ using $\PW$+jets and multijet
events~\cite{1748-0221-7-01-P01001}.

The backgrounds with genuine $\Pgt$ leptons are Drell--Yan $\Pgt\Pgt$, single-top-quark production, dibosons, and the SM $\ttbar$
events in which a $\Pgt$ is produced from a $\PW$ decay.  The $\cPZ$$/\gamma^{\ast}\rightarrow\Pe\Pe$,~$\Pgm\Pgm$ and
$\ttbar\rightarrow\PW^+\cPqb\PW^-\cPaqb$~$\rightarrow \ell^{+}\cPgn\cPqb$$\ell^{-}\cPagn\cPaqb$
events may also contain electrons or muons misidentified as $\Pgt_\mathrm{h}$. The event yields for these backgrounds are estimated from
simulation.

\begin{figure}[htbp]
\begin{center}
\includegraphics[width=0.45\textwidth]{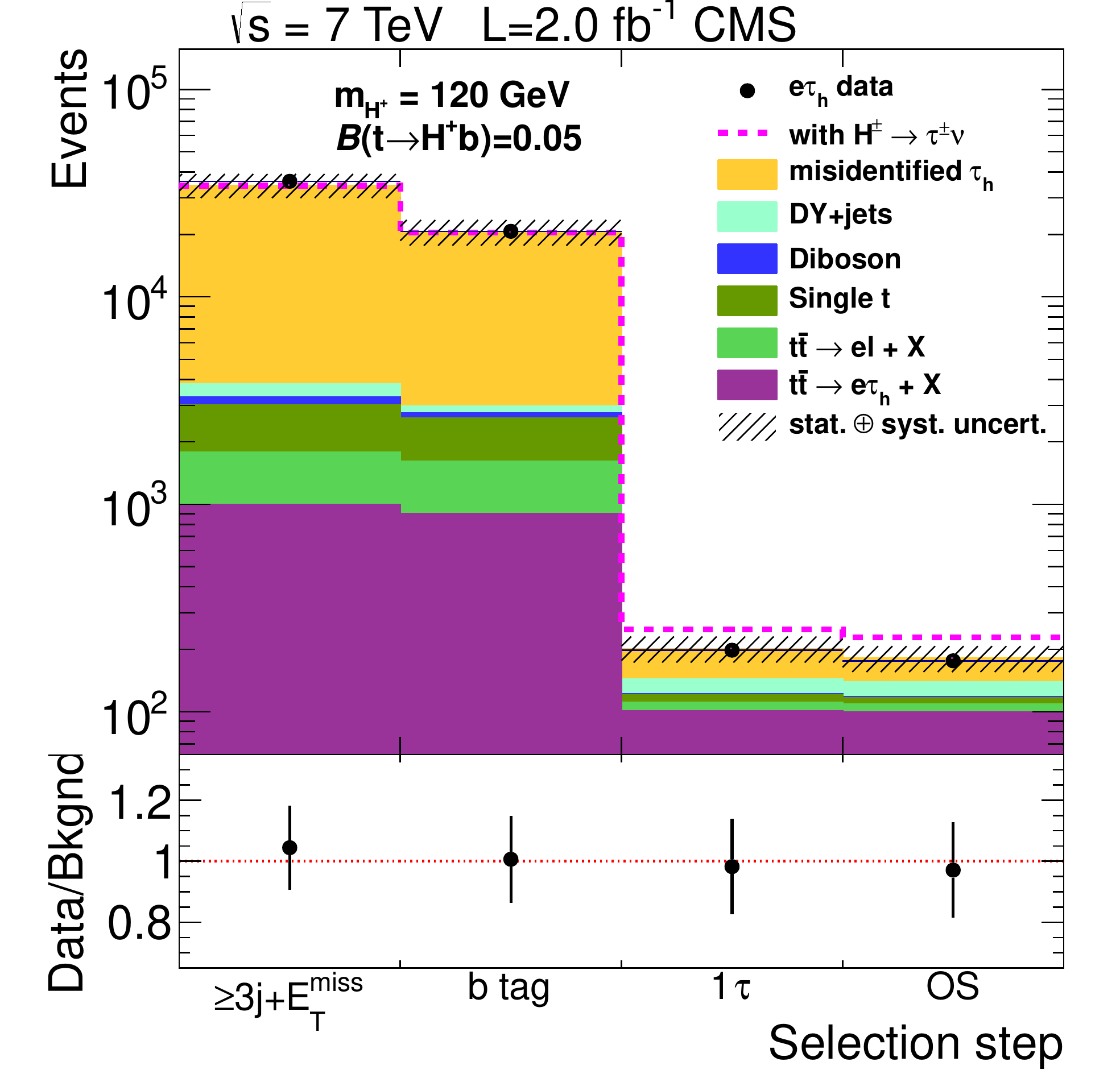}
\includegraphics[width=0.45\textwidth]{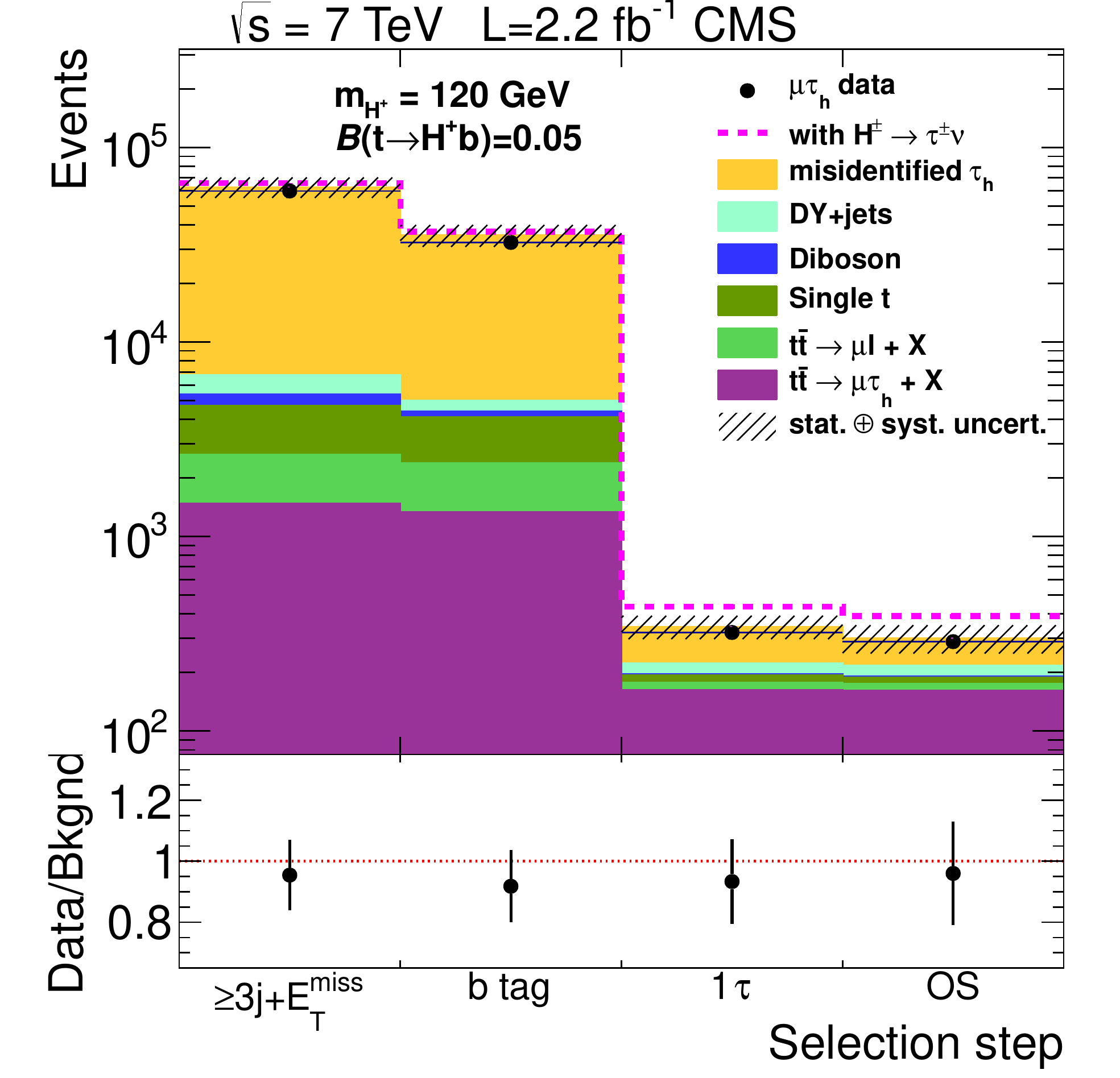}
\caption{The event yields after each selection step for the $\Pe\Pgt_\mathrm{h}$ (left) and $\Pgm\Pgt_\mathrm{h}$ (right) analyses.
         The backgrounds are estimated from simulation and normalized to the standard model prediction.
         The expected event yield in the presence of the $\cPqt \rightarrow \PH^{+}\cPqb$, $\PH^{+}$$\rightarrow\Pgt^{+}\Pgngt$
         decays is shown
         as a dashed line for $m_{\PH^{+}} = 120\GeV$ and under the assumption that $\mathcal{B}(\cPqt \rightarrow\PH^{+}\cPqb) = 0.05$.
	 The bottom panel shows the ratios of data over background with the total uncertainties.
         OS indicates the requirement to have opposite electric charges for a $\Pgt_\mathrm{h}$ and a $\Pe$ or $\Pgm$.
	 Statistical and systematic uncertainties are added in quadrature.}
\label{fig:leptonic_ltau_selections_fig1andfig2}
\end{center}
\end{figure}

The data and the simulated event yield at various stages of event selection, described above, for the $\Pe\Pgt_\mathrm{h}$ ($\Pgm\Pgt_\mathrm{h}$)
analysis are shown in Fig.~\ref{fig:leptonic_ltau_selections_fig1andfig2} left (right). The backgrounds are normalized to the SM
prediction obtained from the simulation. A good agreement is found between data and the SM background.
The expected event yield in the presence of
$\cPqt\rightarrow \PH^{+}\cPqb$, $\PH^{+}$$\rightarrow\Pgt^{+}\Pgngt$ decays is shown as a dashed line for
$m_{\PH^{+}} = 120\GeV$ under the assumption that $\mathcal{B}(\cPqt\rightarrow\PH^{+}\cPqb) = 0.05$.
The observed number of events after the full event selection is shown in Table~\ref{tab:leptonic_ltau_selections_tab1}
along with the expected numbers of events from the various backgrounds, and from the Higgs boson signal processes
$\PW\PH$ and $\PH\PH$ for $m_{\PH^{\pm}} = 120\GeV$. The misidentified $\Pgt$ background measured from the data
is consistent with the expectation from simulation, $42 \pm 4\stat \pm 8\syst$ for the $\Pe\Pgt_{h}$ analysis and
$83 \pm 7\stat \pm 12\syst$ for the $\Pgm\Pgt_{h}$ analysis.
\begin{table*}[htbp]
\begin{center}
\topcaption{Numbers of expected events in the $\Pe\Pgt_\mathrm{h}$ and $\mu\Pgt_\mathrm{h}$ analyses for the backgrounds and the Higgs boson signal
from $\PW\PH$ and $\PH\PH$ processes at $m_{\PH^{+}} = 120\GeV$, and the number of observed events after the final event selection.
Unless stated differently, the expected background events are from simulation.}
\begin{tabular}{|c|c|c|}
\hline
\multicolumn{1}{|c|}{Source\rule{0pt}{12pt}}
& $N_\mathrm{ev}^{\Pe\Pgt_\mathrm{h}} \pm \text{stat.} \pm \text{syst.}$
& $N_\mathrm{ev}^{\Pgm\Pgt_\mathrm{h}} \pm \text{stat.} \pm \text{syst.}$  \\
\hline
\hline
$\PH\PH$+$\PH\PW$, $m_{\PH^{+}}=120\GeV$, $\mathcal{B}(\cPqt\rightarrow\PH^{+}\cPqb)=0.05$ & 51 $\pm$ 3 $\pm$ 8 & 89 $\pm$ 4 $\pm$ 13 \\

\hline
\hline
misidentified $\Pgt$ (from data)                                            &   54 $\pm$ 6 $\pm$ 8 & 89 $\pm$ 9 $\pm$ 11    \\
$\ttbar \rightarrow \PW\cPqb\PW\cPaqb \rightarrow \ell$$\Pgn$$\cPqb$~$\Pgt\Pgn\cPaqb$  &   100 $\pm$ 3 $\pm$ 14  & 162 $\pm$ 4 $\pm$ 23 \\
$\ttbar \rightarrow \PW\cPqb\PW\cPaqb \rightarrow \ell$$\Pgn$$\cPqb$~$\ell\Pgn\cPaqb$  &   9.0 $\pm$ 0.9 $\pm$ 1.8 & 13.0  $\pm$ 1.2  $\pm$ 2.5  \\
$\cPZ$$/\gamma^{\ast}\rightarrow \Pe\Pe,\Pgm\Pgm$                         &   4.8 $\pm$ 1.8 $\pm$ 1.3 & 0.7  $\pm$ 0.7  $\pm$ 0.7   \\
$\cPZ$$/\gamma^{\ast}\rightarrow \Pgt\Pgt$                          &   17.0 $\pm$ 3.3 $\pm$ 3.0 & 26.0  $\pm$ 4.3 $\pm$ 6.1  \\
single top quark                                                &   7.9 $\pm$ 0.4 $\pm$ 1.1 & 13.5  $\pm$ 0.5  $\pm$ 1.9  \\
diboson                                                        &   1.3 $\pm$ 0.1 $\pm$ 0.2 & 2.0  $\pm$ 0.2  $\pm$ 0.3  \\
\hline
Total expected background                                       &   194 $\pm$ 8 $\pm$ 20 & 306 $\pm$ 11 $\pm$ 32  \\
\hline
\hline
Data                                                            &   176 & 288        \\
\hline
\end{tabular}

\label{tab:leptonic_ltau_selections_tab1}
\end{center}
\end{table*}
\section{Analysis of the \texorpdfstring{$\Pe\Pgm$}{e mu} final state \label{sec:emu}}
The event selections are the same as used in the measurement of the top quark pair production cross section
in dilepton final states~\cite{Chatrchyan:2011nb}.

The $\Pe\Pgm$ events are selected by a trigger requiring an electron with $\pt^{\Pe} > 8\GeV$ and a muon with $\pt^{\Pgm} > 17\GeV$;
or an electron with $\pt^{\Pe} > 17\GeV$ and a muon with $\pt^{\Pgm} > 8\GeV$. The amount of data analyzed for this channel
corresponds to an integrated luminosity of $2.27 \pm 0.05\fbinv$.

In the $\Pe\Pgm$ analysis, the events are selected by requiring at least one isolated electron
and at least one isolated muon ($I_\text{rel} < 0.15$) in a cone of radius $\Delta R = 0.3$ around the lepton with
$\pt > 20\GeV$ within $|\eta| < 2.5 \,(2.4)$ for electrons (muons). The event has to have at least two jets with
$\pt > 30\GeV$ within $|\eta| < 2.4$. The leptons are required to be separated from any selected jet by a distance $\Delta R > 0.4$.
The invariant mass of electron-muon pair, $m_{\Pe\Pgm}$, is required to exceed $12\GeV$. The electron and the muon are required to have opposite
electric charges.

The backgrounds considered in the $\Pe\Pgm$ final-state analysis are the following: SM $\ttbar$,
Drell--Yan $\ell \ell$ ($\ell = \Pe, \Pgm, \Pgt$) production in association with jets (DY($\ell\ell$)),
W+jets, single-top-quark production
(dominated by $\cPqt\PW$) and diboson ($\PW\PW$, $\PW\cPZ$, $\cPZ\cPZ$) production. Background yields are all estimated from simulation.
After the signal selection requirements are applied, 95\% of the remaining background is due to SM $\ttbar$ decays.

\begin{figure}[htb]
\begin{center}
\includegraphics[width=0.60\textwidth]{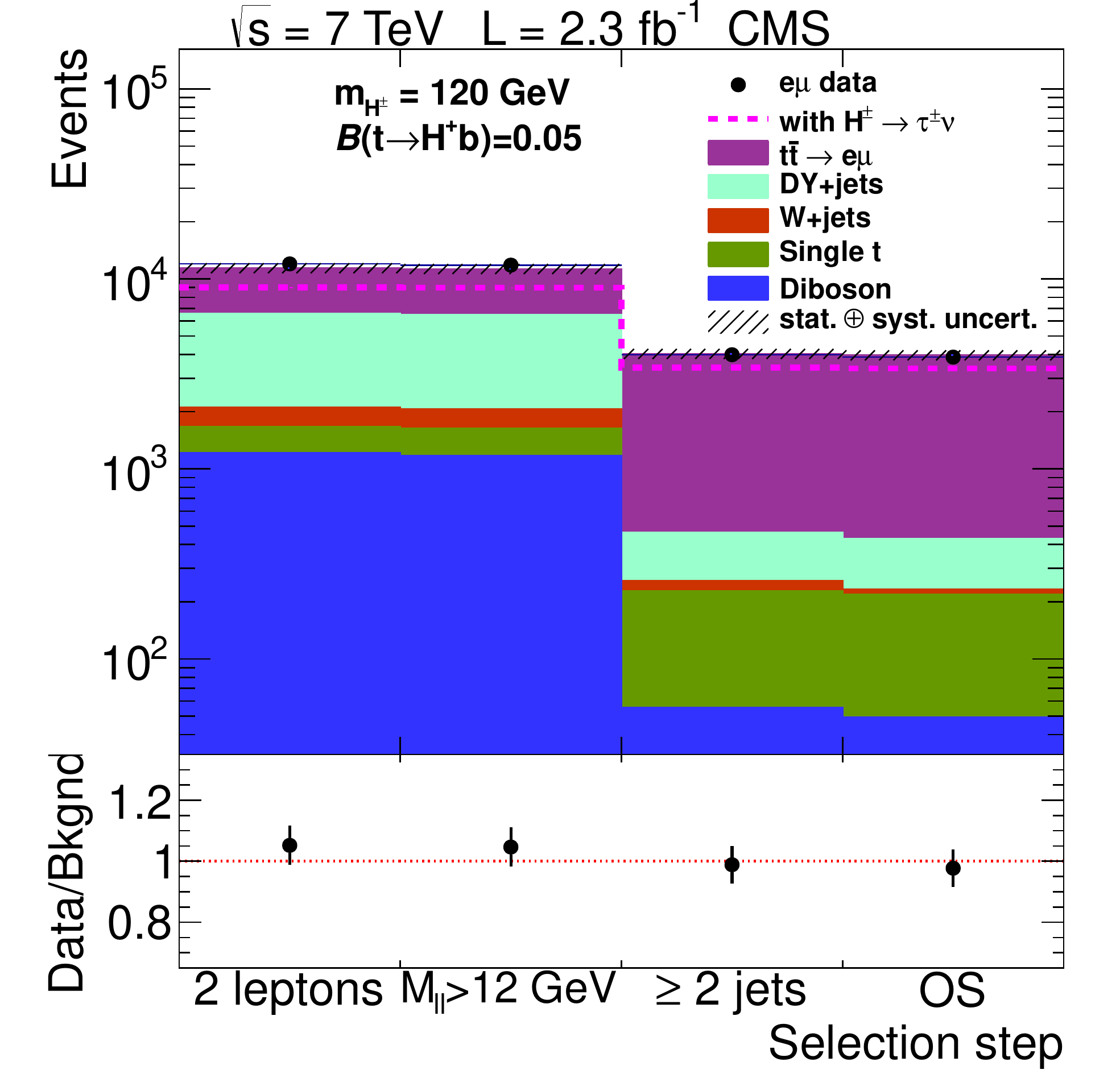}
\caption{The event yield after each selection step for the $\Pe\Pgm$ analysis. The backgrounds are from simulation and
         normalized to the standard model prediction. The expected event yield in the presence of the
         $\cPqt\rightarrow\PH^{+}\cPqb$, $\PH^{+}\rightarrow\Pgt^{+}\Pgngt$ decays is shown as a dashed line for
         $m_{\PH^{+}}= 120\GeV$ under the assumption that $\mathcal{B}(\cPqt\rightarrow\PH^{+}\cPqb) = 0.05$.
	 The bottom panel shows the ratios of data over background with the total uncertainties.
         The requirement for the $\Pe$ and $\Pgm$ to have opposite electric charges is labelled as OS.
	 Statistical and systematic uncertainties are added in quadrature.}
\label{fig:leptonic_emu_selections_fig6}
\end{center}
\end{figure}

The data and simulated event yields at various stages of the event selection are shown in Fig.~\ref{fig:leptonic_emu_selections_fig6}.
The backgrounds are normalized to the standard model prediction obtained by simulation.
A good agreement between the data and the standard model expectations is found.
The expected event yield in the presence of $\cPqt\rightarrow\PH^{+}\cPqb$, $\PH^{+}\rightarrow\Pgt^{+}\Pgngt$ decays is
shown as a dashed line for $m_{\PH^{+}} = 120\GeV$ under the assumption that $\mathcal{B}(\cPqt\rightarrow\PH^{+}\cPqb) = 0.05$.
It is smaller than the expectation from the SM alone ($\mathcal{B}(\cPqt\rightarrow\PH^{+}\cPqb) = 0$) because the selection efficiency is smaller
for $\PH^{+}\rightarrow\Pgt^+\Pgngt\rightarrow\ell^+\Pgn_{\ell}\Pagngt\Pgngt$ than for
$\PW^{+}\rightarrow\ell^+\Pgn_{\ell}$ decay owing to the softer lepton $\pt$ spectrum.

The numbers of expected events for the backgrounds and the Higgs boson signal processes from $\PW\PH$ and $\PH\PH$ modes at
$m_{\PH^{\pm}} = 120\GeV$, and the number of observed events after all selection requirements are summarized in
Table~\ref{tab:leptonic_emu_selections_tab1}.
\begin{table*}[htbp]
\begin{center}
\topcaption{Number of expected events in the $\Pe\Pgm$ analysis for the backgrounds, the Higgs boson signal from
$\PH\PH$ and $\PW\PH$ processes at $m_{\PH ^{+}}=120\GeV$, and the number of observed events after all selection requirements.
The expected background events are from simulation.}
\setlength{\extrarowheight}{1pt}
\begin{tabular}{|c|c|}
\hline
 Source\rule{0pt}{12pt}                                 & $N_\mathrm{ev}^{\Pe\mu} \pm \text{stat.} \pm \text{syst.}$              \\
\hline
\hline
$\PH\PH$+$\PW\PH$, $m_{\PH^{+}} = 120\GeV$, $\mathcal{B}(\cPqt\rightarrow\PH^{+}\cPqb) = 0.05$       & 125 $\pm$ 9 $\pm$ 13  \\
\hline
\hline
$\ttbar$ dileptons 				& 3423 $\pm$ 35 $\pm$ 405    \\
other $\ttbar$     				& 23 $\pm$ 3 $\pm$ 3    \\
$\cPZ$$/\gamma^{\ast}\rightarrow \ell\ell$ 		& 192 $\pm$ 12 $\pm$ 19       \\
$\PW$+jets 						& 14 $\pm$ 6 $\pm$ 2         \\
single top quark 				& 166 $\pm$ 3 $\pm$ 18         \\
diboson 				        & 48 $\pm$ 2 $\pm$ 5    \\
\hline
Total expected background		        	& 3866 $\pm$ 38 $\pm$ 406     \\
\hline
\hline
Data						& 3875	            \\
\hline
\end{tabular}

\label{tab:leptonic_emu_selections_tab1}
\end{center}
\end{table*}

\section{Systematic uncertainties \label{sec:systematics}}
The sources and the size of the systematic uncertainties are listed in Tables~\ref{tab:fullyhadronic_systematics_tab1},
~\ref{tab:leptonic_ltau_systematics_tab1}, and ~\ref{tab:leptonic_emu_systematics_tab1}. In all of the analyses
the following effects are taken into account:

\begin{itemize}
\item{the uncertainty on the jet energy scale (JES), jet energy resolution (JER), and $\ETmiss$ scale.
      This uncertainty is estimated following the procedure outlined in Ref.~\cite{jme10}; an uncertainty of 3\% on the $\Pgt_\mathrm{h}$
      energy scale is included;}
\item{the theoretical uncertainties on the signal and background cross sections;}
\item{the uncertainty on pileup modelling due to the reweighting of simulated events according to the measured distribution
      of the number of vertices;}
\item{the uncertainty due to the limited number of events available in the simulated samples (MC stat.);}
\item{an estimated 2.2\% uncertainty in the integrated luminosity~\cite{smp-12-008}.}
\end{itemize}

In addition, for the fully hadronic channel the following systematic uncertainties are taken into account:
\begin{itemize}
\item{the uncertainty on trigger efficiencies. The efficiency of the $\Pgt$ part of the trigger is evaluated using $\cPZ \rightarrow$~$\Pgt\Pgt$
      events. It is used for the ``EWK+$\ttbar$~$\Pgt$" background estimate. The data-to-simulation correction factor for the trigger on
      $\ETmiss$ is evaluated using $\ttbar$ events with an uncertainty estimated to be ${\simeq}10\%$. The data-to-simulation correction
      factors for the efficiency of the trigger on $\Pgt_\mathrm{h}$ and on $\ETmiss$ are used for the $\PW\PH$, $\PH\PH$ signal and
      ``EWK+$\ttbar$~no-$\Pgt$" background estimates;}
\item{the uncertainty on the estimate of the multijet background from data;}
\item{the uncertainty on the estimate of ``EWK+$\ttbar$~$\Pgt$" background due to the uncertainty on the $\Pgt_\mathrm{h}$ jet energy scale,
      the selection of muons in the control sample, the limited number of events in the control sample, the contamination
      from multijet background, and the fraction of $\PW$~$\rightarrow\Pgt\rightarrow\Pgm$ events ($f_{\PW}$$_{\rightarrow\Pgt\rightarrow\Pgm}$)
      in the control sample;}
\item{the uncertainty in the application of the lepton veto. It is estimated from the uncertainty in the lepton reconstruction, identification,
      and isolation efficiencies of 2\% (1\%) for electrons (muons), which is measured using $\cPZ$~$\rightarrow \ell \ell$ ($\ell = \Pe,\Pgm$) events;}
\end{itemize}

In addition, for the analyses with $\Pgt_\mathrm{h}$ in the final state ($\Pgt_\mathrm{h}$+jets, $\Pe\Pgt_\mathrm{h}$, $\mu\Pgt_\mathrm{h}$),
the following systematic uncertainties are taken into account:
\begin{itemize}
\item{the uncertainty on the efficiency of $\Pgt$ identification, estimated to be 6\%~\cite{1748-0221-7-01-P01001};}
\item{the uncertainty on the rate of misidentification of a jet as a $\Pgt_\mathrm{h}$ or of a lepton as a $\Pgt_\mathrm{h}$,
      each estimated to be 15\%~\cite{1748-0221-7-01-P01001};}
\item{the uncertainty on the efficiency of b tagging, $5.4\%$~\cite{btag-11-001};}
\item{the uncertainty on the rate of misidentification of a jet as a b quark, $10\%$~\cite{btag-11-001};}
\end{itemize}

In the $\Pe\Pgt_\mathrm{h}$ and $\Pgm\Pgt_\mathrm{h}$ analyses the uncertainty in the estimation of the misidentified $\Pgt$ background has two sources:
the limited number of events for the measurement of the $\Pgt$ misidentification rate and the difference in the $\Pgt$
misidentification rates for jets originating from a quark with respect to jets originating from a gluon.

Finally the uncertainty on the reconstruction, identification, and isolation efficiency of an electron or a muon is taken into account in
the $\Pe\Pgt_\mathrm{h}$, $\Pgm\Pgt_\mathrm{h}$, and $\Pe\Pgm$ analyses. It is estimated to be ${\simeq}2$--3\%.

The full sets of systematic uncertainties are used as input to the exclusion limit calculation.
\begin{table*}[htbp]
\begin{center}
\topcaption{The systematic uncertainties on event yields (in percent) for the $\Pgt_\mathrm{h}$+jets analysis for background processes
and for the Higgs boson signal processes $\PW\PH$ and $\PH\PH$ in the range of $m_{\PH^{+}} = 80$--160\GeV.
The range of errors for the signal processes is given for the Higgs boson mass range of 80--160\GeV.}
\footnotesize{
\setlength{\extrarowheight}{1.5pt}
\begin{tabular}{|l|c|c|c|c|c|c|c|c|c|c|}
  \hline
                                   & $\PH\PH$   &  $\PW\PH$  & multi-  & \multicolumn{3}{c|}{EWK+$\ttbar$~$\Pgt$}
& \multicolumn{3}{c|}{EWK+$\ttbar$~no-$\Pgt$}
\\\cline{5-10}
                                   &      &      & jets   & Emb.data & Res.DY & Res.WW & $\ttbar$ & $\cPqt\PW$  & $\PW$+jets \\
  \hline
  \hline
    JES+JER+$\ETmiss$    &4.7--14   & 9.0--18 &    & 6.6 & 26 & 23 & 8.1 & 2.4  & $<$10   \\
\hline
    cross section                  &$^{+7.0}_{-10.0}$&$^{+7.0}_{-10.0}$&  & & 4.0 & 4.0 & $^{+7.0}_{-10.0}$ &  8.0 &  5.0      \\
\hline
    pileup modeling               & 0.3--4.2& 0.6--5.2&    &     & 7.6& 3.9& 7.1 & 15  & 10   \\
\hline
    MC stat                       & 6.2--11 & 7.0--10 &    &     & 29 & 66 & 28  & 49  & 71    \\
\hline
    luminosity                     & \multicolumn{2}{c|}{ 2.2}   &    &    & \multicolumn{5}{c|}{ 2.2} \\
\hline
    trigger                        & 12--13  & 13      &    & 11  & 12 & 11 & 12  & 11 &  14  \\
\hline
    multijet stat.                 &         &         &6.5 &     &    &    &     &     &           \\
\hline
    multijets syst.                &         &         &3.8 &     &    &    &     &     &           \\
\hline
    $\mu$ sample stat.             &         &         &    & 3.4  &    &    &     &     &        \\
\hline
    multijet contamin.             &         &         &    & 0.3 &    &    &     &     &        \\
\hline
 $f_{{\PW}\to\Pgt\to\Pgm}$              &         &         &    & 0.7 & 0.1& 0.1&     &     &        \\
\hline
    muon selections                &         &         &    & 0.5 & 0.1& 0.1&     &     &        \\
\hline
    lepton veto                    &0.3--0.5 & 0.5--0.7&    &     & 0.9& 1.2& 0.9 & 0.6 & 0.3     \\
\hline
    $\Pgt$-jet id                  &  6.0  & 6.0     &    & 6.0 & 6.0& 6.0&     &     &    \\
\hline
jet,~$\ell \rightarrow \Pgt$ misident. &         &         &    &     &    &    &  \multicolumn{3}{c|}{15}  \\
\hline
    b-jet tagging                  &1.1--2.1 & 1.0-1.7 &    &     &    &    & 1.4 & 1.6 &           \\
\hline
    jet$\rightarrow\cPqb$ misident.    &         &         &    &     & 2.0& 2.6&     &     & 4.8 \\
\hline
\end{tabular}
}
\label{tab:fullyhadronic_systematics_tab1}
\end{center}
\end{table*}

\begin{table*}[htpb]
\begin{center}
\topcaption{The systematic uncertainties on event yields (in percent) for the $\Pgm\Pgt_\mathrm{h}$ analysis for the background processes
and for the Higgs boson signal processes $\PW\PH$ and $\PH\PH$ for $m_{\PH^{+}} = 120\GeV$.}
\footnotesize{
\setlength{\extrarowheight}{1.5pt}
\begin{tabular}{|l|c|c|c|c|c|c|c|c|c|}
  \hline
                                   & $\PH\PH$   &  $\PW\PH$  & $\ttbar$$_{\ell \Pgt}$ & $\ttbar$$_{\ell \ell}$ &misident. $\Pgt$ & Single top & diboson & DY($\Pgm\Pgm$) & DY($\Pgt\Pgt$) \\
  \hline
  \hline
    JES+JER+$\ETmiss$     &  6.0 & 5.0  &      5.0               &        4.0              &             &  6.0       &11.0&    100.0         &    22.0     \\
\hline
    cross section                  &      \multicolumn{4}{c|}{$^{+7.0}_{-10}$}                        &             &   8.0      & 4.0& \multicolumn{2}{c|}{4.0}        \\
\hline
    pileup modeling                & 4.0  & 2.0  &      2.0               &        8.0              &             &  2.0       &3.0 &      25.0       &    4.0       \\
\hline
    MC stat                       &  5.0 &  4.0 &      2.0               &       9.0               &             &   4.0      & 9.0&    100.0         &   16.0         \\
\hline
    luminosity                     &  \multicolumn{4}{c|}{ 2.2 } &          & \multicolumn{4}{c|}{ 2.2 }                                                   \\
\hline
    $\Pgt$-jet id                  & 6.0  &  6.0 &         6.0            &                         &             &  6.0       & 6.0&                 &     6.0      \\
\hline
jet,~$\ell \rightarrow \Pgt$ misident.&      &      &                        &        15.0             &             &            &    &      15.0       &             \\
\hline
    b-jet tagging                  & 6.0  & 5.0  &      5.0               &        5.0              &             &  7.0       &    &                 &             \\
\hline
    jet$\rightarrow \cPqb$ misident.       &      &      &                        &                         &             &            &8.0 &      8.0        &    9.0       \\
\hline
   misident.\ $\Pgt$ (stat.)          &      &      &                        &                         &   10.0      &            &    &                 &              \\
\hline
   misident.\ $\Pgt$ (syst.)          &      &      &                        &                         &   12.0      &            &    &                 &              \\
\hline
    lepton selections              &     \multicolumn{4}{c|}{ 2.0}         &                        &   \multicolumn{4}{c|}{ 2.0}                                   \\
\hline
\end{tabular}

}
\label{tab:leptonic_ltau_systematics_tab1}
\end{center}
\end{table*}

\begin{table*}[htbp]
\begin{center}
\topcaption{The systematic uncertainties on event yields (in percent) for the $\Pe\Pgm$ analysis for the background processes
and for the Higgs boson signal processes $\PW\PH$ and $\PH\PH$ at $m_{\PH^{+}} = 120\GeV$.}
\begin{tabular}{|l|c|c|c|c|c|c|c|}
  \hline
                                   & $\PH\PH$   &  $\PW\PH$     & $\ttbar$ & DY($\ell\ell$) & W+jets & Single top & diboson   \\
  \hline
  \hline
    JES+JER+$\ETmiss$     & 2.1   &  2.0   &    2.0     &     6.0          &   10.8    &     4.0   & 6.5     \\
  \hline
    cross section                  & \multicolumn{3}{c|}{$^{+7}_{-10}$} &      4.3  &   5.0   &     7.4   & 4.0    \\
\hline
    pileup modeling               & 4.5   & 4.5    &   5.0      &     5.5          &   4.0   &     5.5   & 5.5     \\
\hline
    MC stat               & 5.3   & 7.9    &   1.0      &      6.5         &   42.9  &     1.9   & 4.3     \\
\hline
    luminosity                     &                      \multicolumn{7}{c|}{2.2}                                            \\
\hline
   dilepton selection              &                      \multicolumn{7}{c|}{2.5} \\
\hline
\end{tabular}

\label{tab:leptonic_emu_systematics_tab1}
\end{center}
\end{table*}

In the $\Pgt$+jets analysis the $\mt$  distribution shown in Fig.~\ref{fig:fullyhadronic_selections_fig2} is used
in a binned maximum-likelihood fit in order to extract a possible signal. Other channels use event counting only
for setting the limits. The uncertainties on the shapes for the multijet and ``EWK+$\ttbar$~$\Pgt$" backgrounds
derived from data are evaluated taking account of the corresponding uncertainty in every bin of the $\mt$  distribution.
In addition, the $\mt$  shape uncertainty for the ``EWK+$\ttbar$~$\Pgt$" background, related to the
$\Pgt_\mathrm{h}$ energy scale uncertainty, is taken into account in the fit. For the signal and the small ``EWK+$\ttbar$~no-$\Pgt$" background
the $\mt$  shape uncertainty in the JES+JER+$\ETmiss$ scale is evaluated from simulation.

\section{Evaluation of limits on \texorpdfstring{$\mathcal{B}(\cPqt\rightarrow\PH^{+}\cPqb)$}{B( t to Hb)} \label{sec:comblimit}}

\begin{figure}[htbp]
\includegraphics[width=0.45\textwidth]{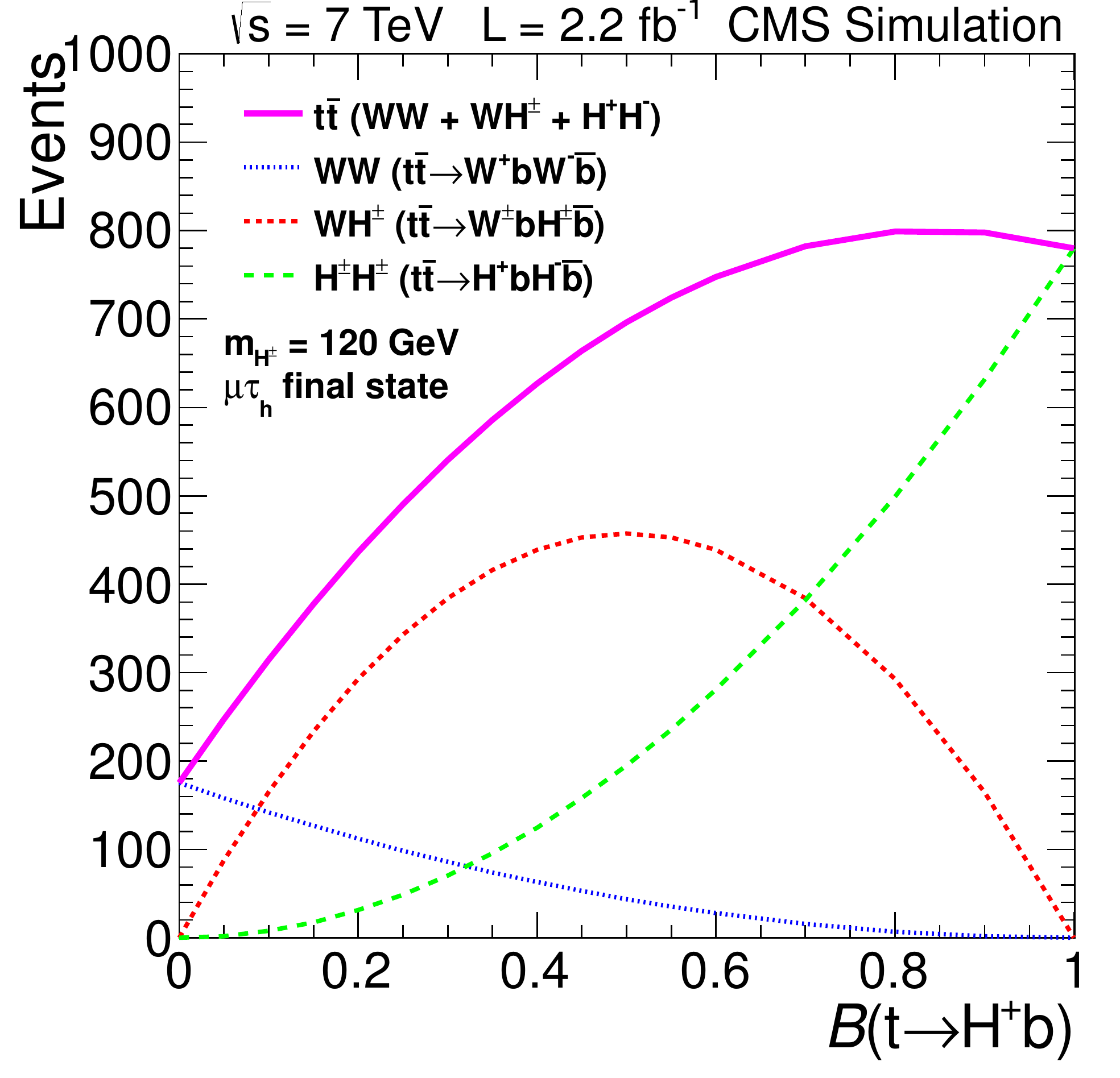}
\includegraphics[width=0.45\textwidth]{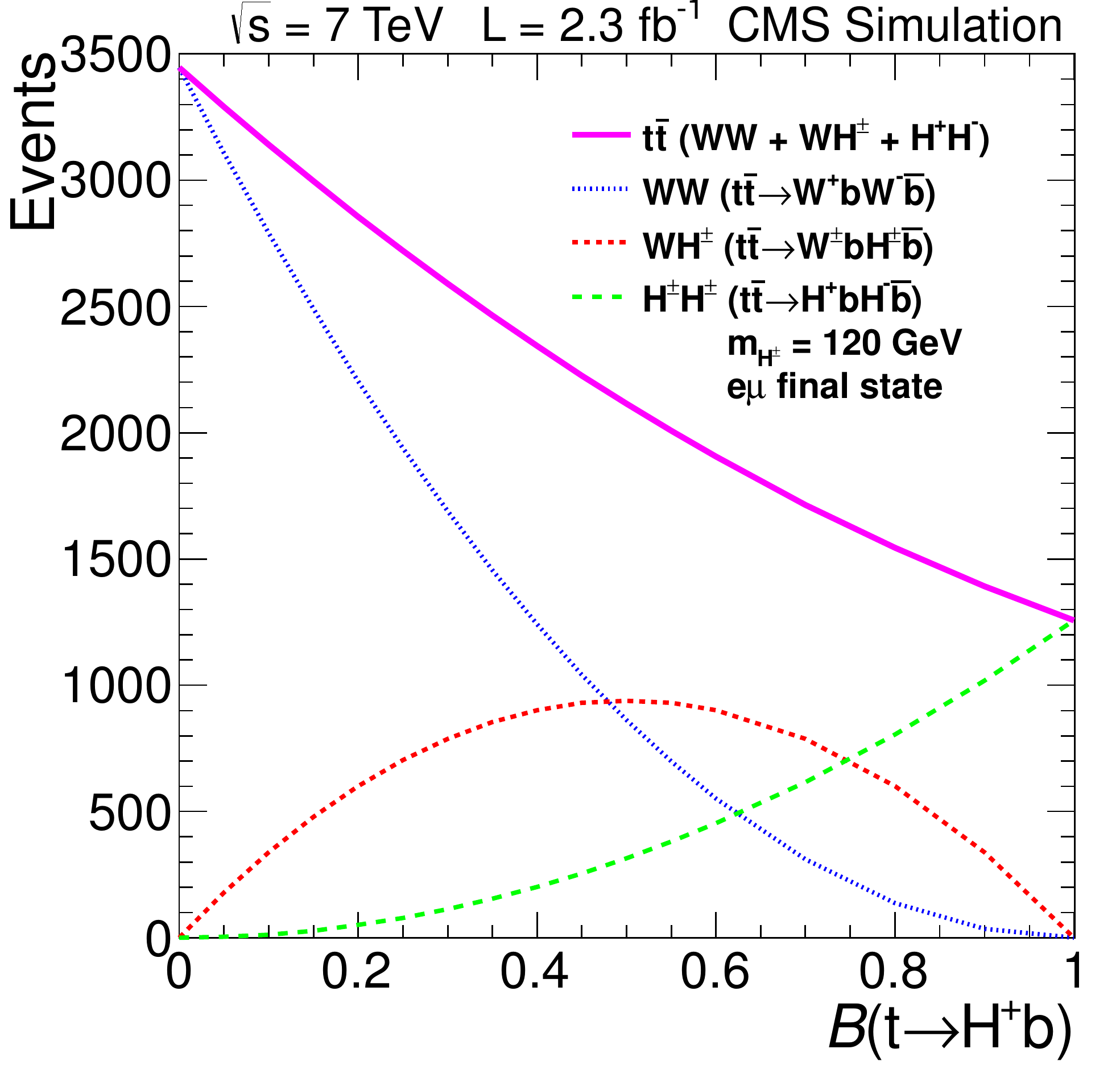}
\caption{The expected number of $\ttbar$ events after event selection for the $\Pgm \Pgt_\mathrm{h}$ (left) and $\Pe\Pgm$ (right)
         final states as a function of the branching fraction $\mathcal{B}(\cPqt\rightarrow\PH^{+}\cPqb)$ for $m_{\PH^{+}} = 120\GeV$.
         Expectations are shown separately for the $\PW\PH$, $\PH\PH$, and $\PW\PW$ contributions.}
\label{fig:limitcombination_fig1and2}

\end{figure}

The expected number of $\ttbar$ events, after final event selection, is shown in Fig.~\ref{fig:limitcombination_fig1and2}
for the $\Pgm\Pgt_\mathrm{h}$ (left) and $\Pe\Pgm$ (right) analyses as a function of the branching fraction $\mathcal{B}(\cPqt\rightarrow\PH^{+}\cPqb)$
for $m_{\PH^{+}} = 120\GeV$. Expectations are shown separately for contributions from $\PW\PH$, $\PH\PH$, and
$\ttbar \rightarrow \PW\cPqb\PW\cPaqb$ ($\PW\PW$) processes. In the e$\Pgt_\mathrm{h}$, $\Pgm\Pgt_\mathrm{h}$, and fully hadronic analyses the total
$\ttbar$ event yield
($N_{\ttbar}^\mathrm{MSSM}$) from $\PW\PW$, $\PW\PH$, and $\PH\PH$ processes is larger than the yield from the standard model
$\ttbar \rightarrow \PW\cPqb\PW\cPaqb$ process ($N_{\ttbar}^\mathrm{SM}$). This is due to the fact that the branching fraction for
the Higgs boson decay into $\Pgt\Pgngt$ is larger than the corresponding branching fraction for $\PW$ boson decay.
For the $\Pe\Pgm$ analysis the total $\ttbar$ event yield is smaller than that expected from the standard model.

Assuming that any excess or deficit of events in data, when compared with the expected background contribution, is due to the $\cPqt \rightarrow \PH^{+}\cPqb, ~ \PH^{+}$$\rightarrow \Pgt^{+} \Pgngt$ decays, the value
of $x = \mathcal{B}(\cPqt \rightarrow \PH^{+}\cPqb)$ for each individual analysis can be related to the difference $\Delta N$ between the
observed number of events and the predicted background contribution through the following equation:

\begin{equation}
\label{eq:limit1}
\Delta N = N_{\cPqt\cPaqt}^{\mathrm{MSSM}} - N_{\ttbar}^{\mathrm{SM}} =
2x(1-x) N_{\PW\PH} + x^2 N_{\PH\PH} + [(1-x)^{2} - 1] N_{\ttbar}^{\mathrm{SM}} .
\end{equation}

In this equation $N_{\PW\PH}$ is estimated from simulation forcing the first top quark to decay to $\PHpm\cPqb$ and the
second to $\PW^{\mp}\cPqb$, and $N_{\PH\PH}$ forcing both top quarks to decay to $\PHpm\cPqb$.  In the $\Pe\Pgt_\mathrm{h}$, $\Pgm\Pgt_\mathrm{h}$,
and $\Pe\Pgm$ analyses, $N_{\ttbar}^\mathrm{SM}$ is evaluated from simulation, as given by the $\ttbar$ background in
Table~\ref{tab:leptonic_ltau_selections_tab1} and~\ref{tab:leptonic_emu_selections_tab1}. In the $\Pgt_\mathrm{h}$+jets analysis,
most of the $\ttbar \rightarrow \PW\cPqb\PW\cPaqb$ yield is derived directly from data, so it does not contribute to $\Delta N$
whatever the value of $x$. In other words if an $\PH^{+}$ SUSY signal is present in the data, affecting the
$\ttbar \rightarrow \PW\cPqb\PW \cPaqb$ rate, it also affects the data driven background estimate for this rate and therefore this contribution
disappears in the difference $\text{data} - \text{background}$.
In this case $N_{\ttbar}^\mathrm{SM}$ contains only the small $\ttbar$ contribution
included in the ``EWK+$\ttbar$ no-$\Pgt$" background in Table~\ref{tab:fullyhadronic_selections_tab1}, which is derived from
simulation: $N_{\ttbar}^\mathrm{SM} = 2.1 \pm 0.6\stat \pm 0.5\syst$.

The CLs method~\cite{CLs,Junk:1999kv} is used to obtain an upper limit, at 95\% confidence level (CL),
on $x = \mathcal{B}(\cPqt \rightarrow \PH^{+}\cPqb)$ using Eq.~\ref{eq:limit1} for each final-state analysis and for their combination.
The background and signal uncertainties described in Section~\ref{sec:systematics} are modeled with a log-normal probability
distribution function and their correlations are taken into account. In the $\Pgt$+jets analysis the $m_\mathrm{T}$ distribution shown in
Fig.~\ref{fig:fullyhadronic_selections_fig2} is used in a binned maximum-likelihood fit in order to extract a possible signal.
For the $\Pe\Pgt_\mathrm{h}$, $\Pgm\Pgt_\mathrm{h}$, and $\Pe\Pgm$ final states only event counting is used to obtain the upper limits.

\begin{figure}[htbp]
\begin{center}
\includegraphics[width=0.45\textwidth]{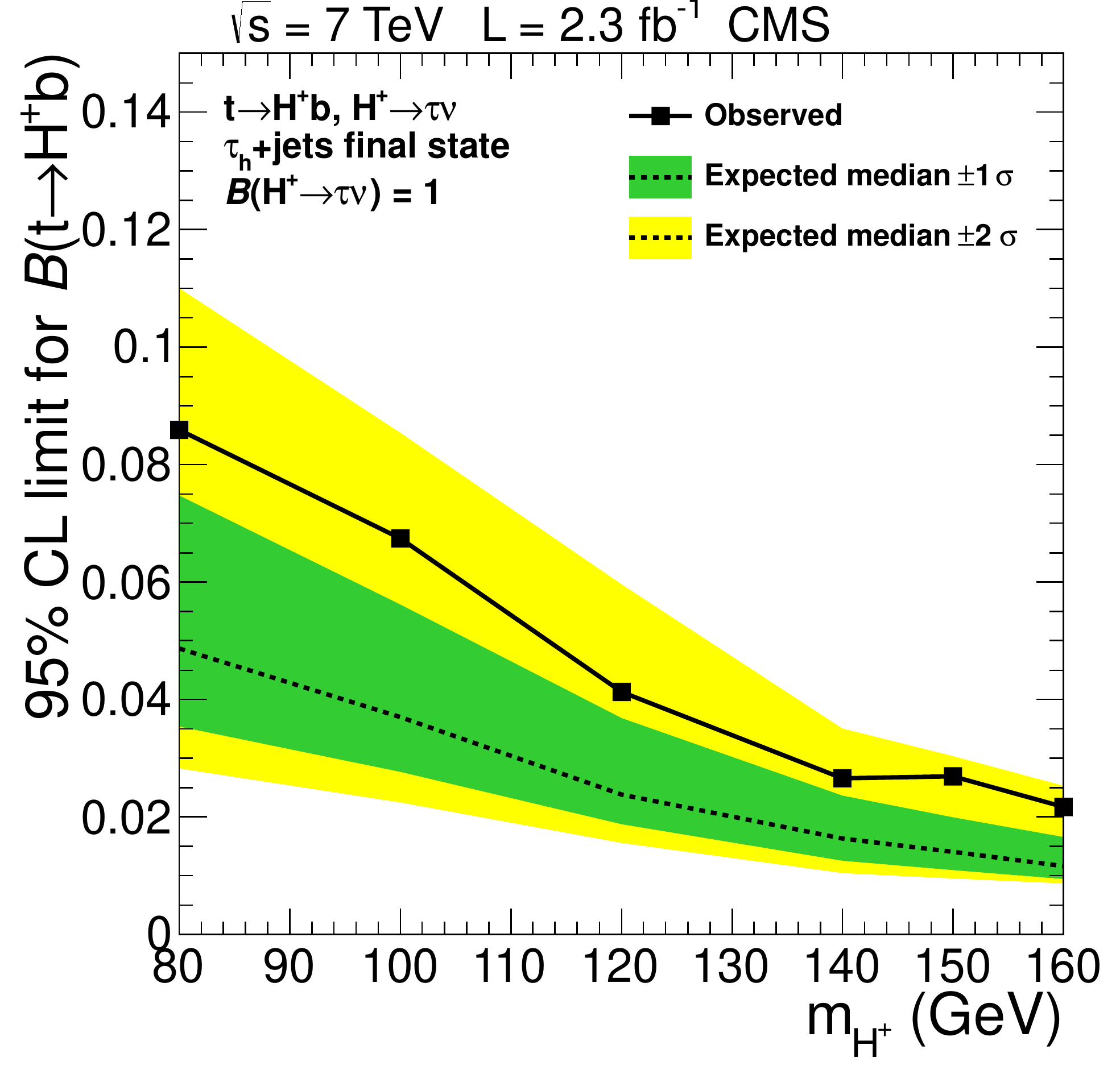}
\includegraphics[width=0.45\textwidth]{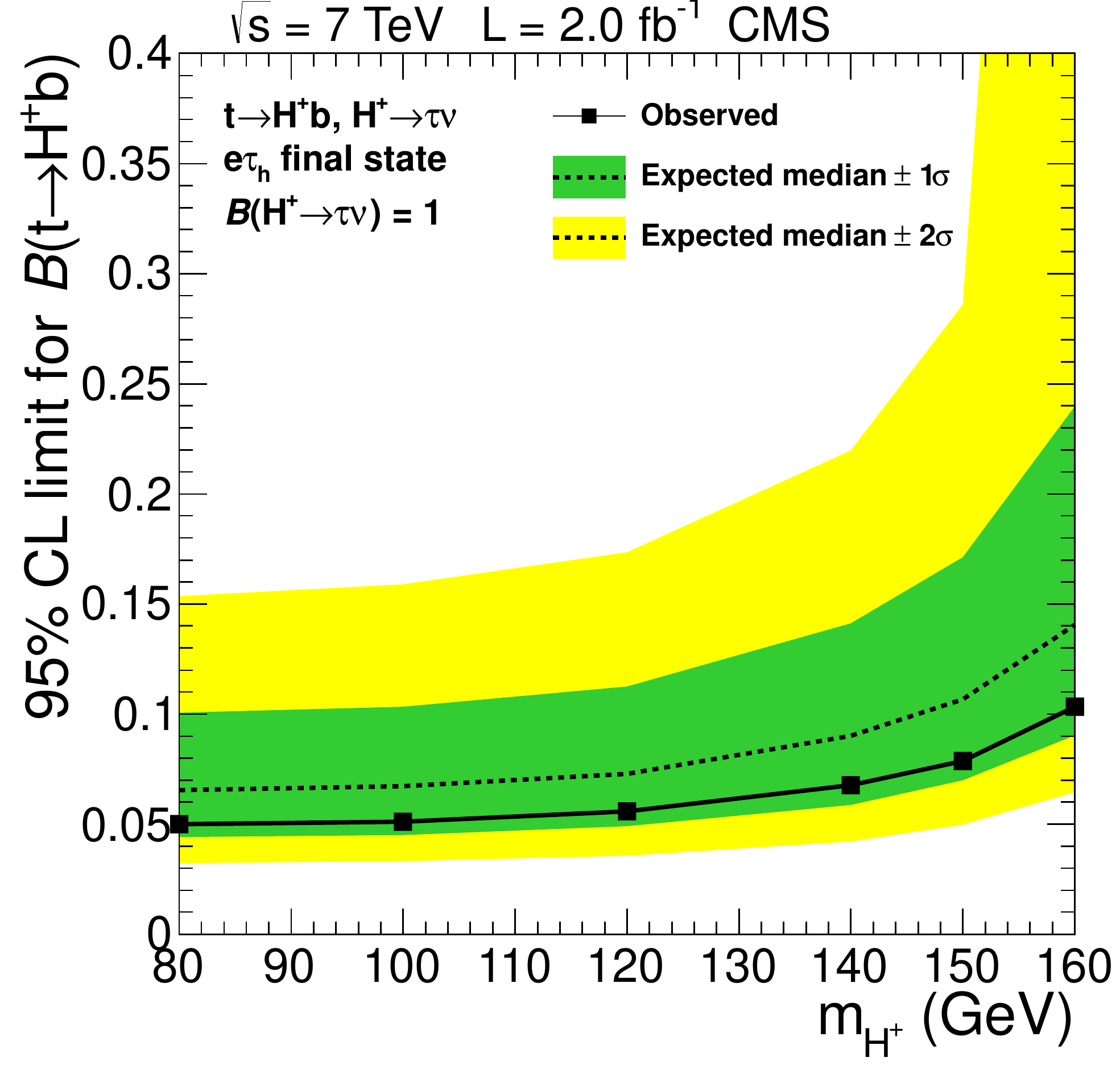}
\caption{Upper limit on $\mathcal{B}(\cPqt \rightarrow \PH^{+}\cPqb)$ as a function of $m_{\PH^{+}}$
         for the fully hadronic (left) and the e$\Pgt_\mathrm{h}$ (right) final states. The $\pm 1 \sigma$ and $\pm 2 \sigma$ bands
         around the expected limit are also shown.}
\label{fig:limitcombination_fig3and4}
\end{center}
\end{figure}

\begin{figure}[htbp]
\begin{center}
\includegraphics[width=0.45\textwidth]{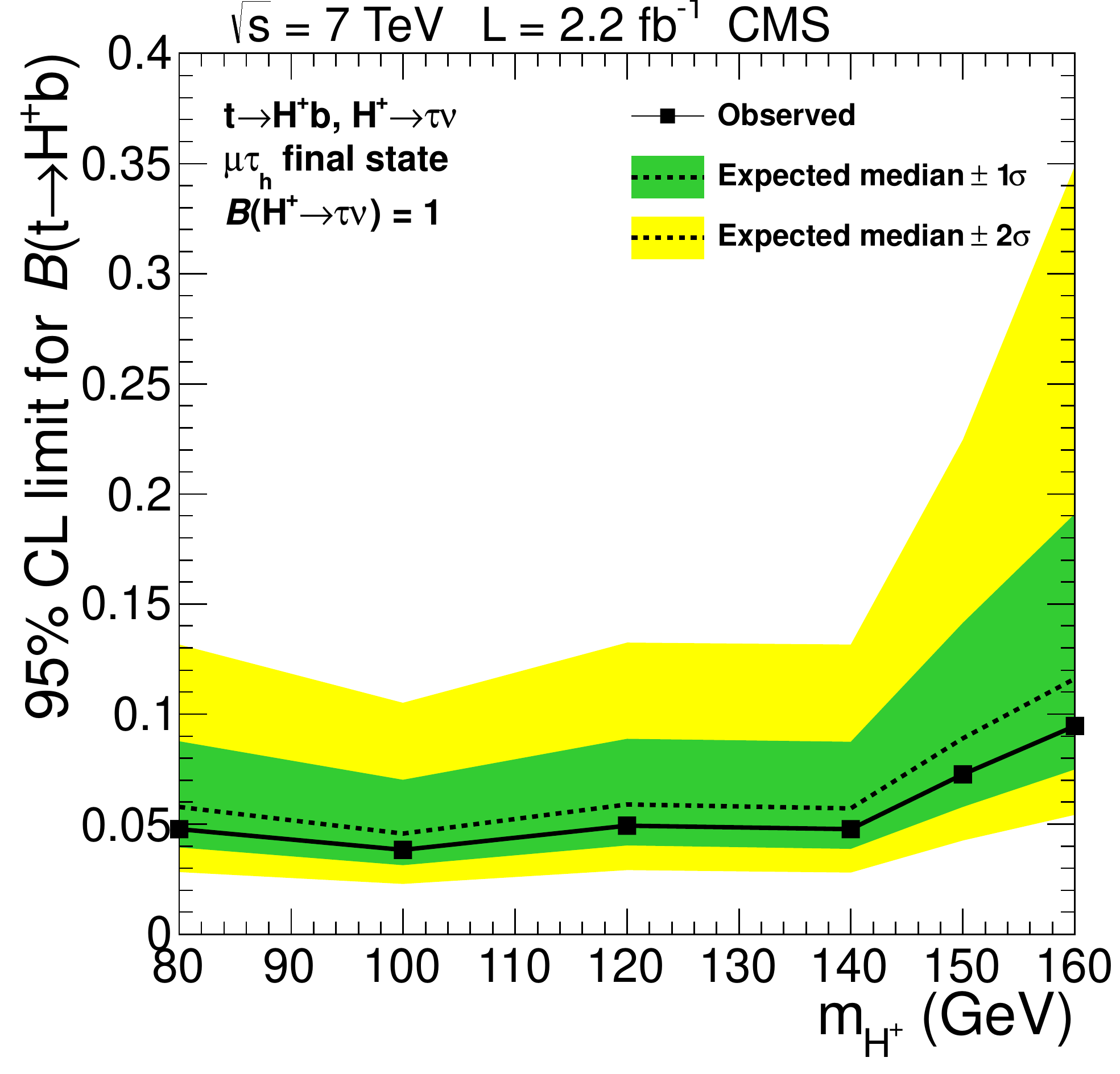}
\includegraphics[width=0.45\textwidth]{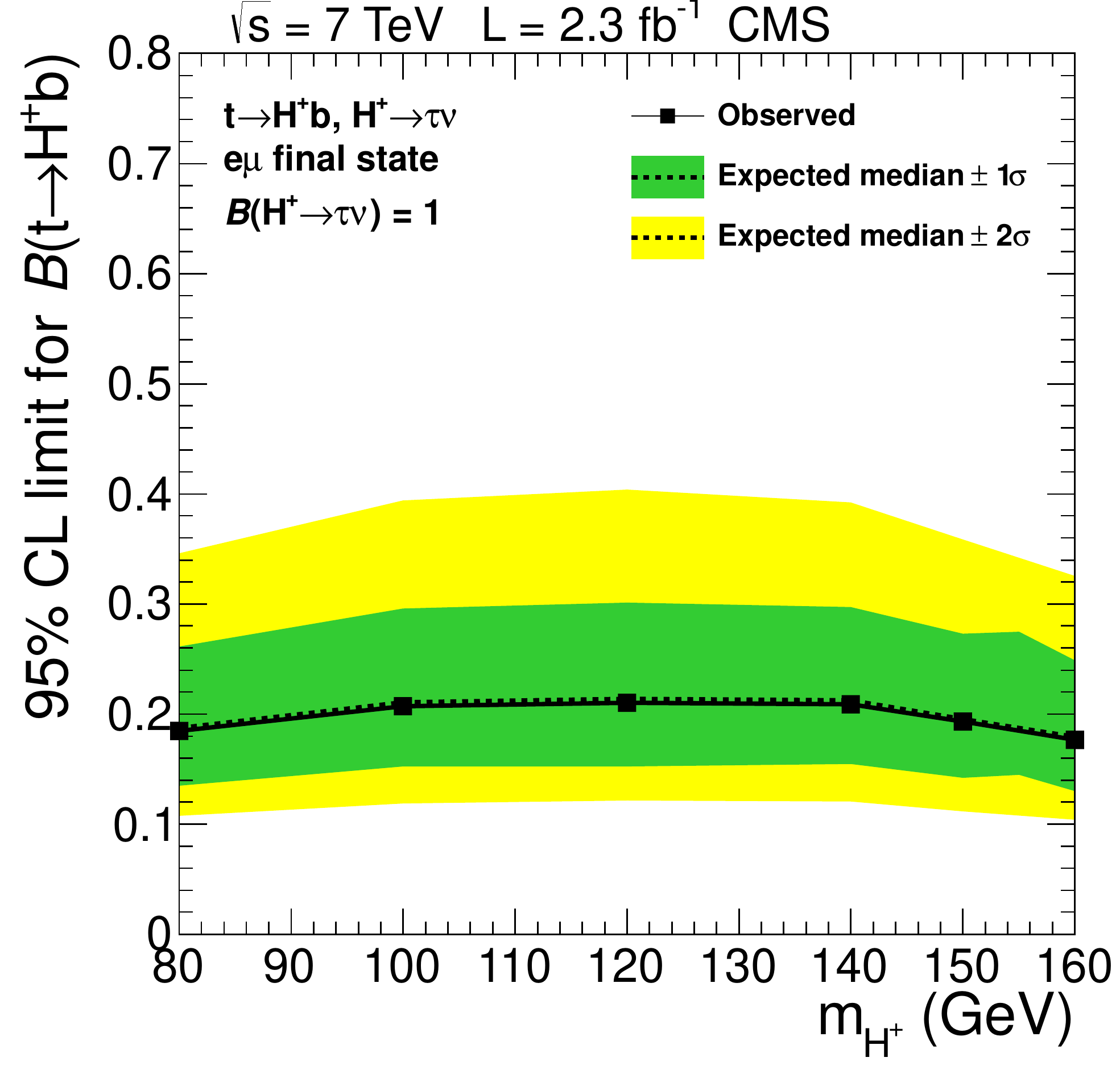}
\caption{Upper limit on $\mathcal{B}(\cPqt\rightarrow \PH^{+}\cPqb)$ as a function of $m_{\PH^{+}}$
         for the $\Pgm\Pgt_\mathrm{h}$ (left) and $\Pe\Pgm$ (right) final states. The $\pm 1 \sigma$ and $\pm 2 \sigma$ bands
         around the expected limit are also shown.}

\label{fig:limitcombination_fig5and6}
\end{center}
\end{figure}

\begin{figure}[htbp]
\begin{center}
\includegraphics[width=0.45\textwidth]{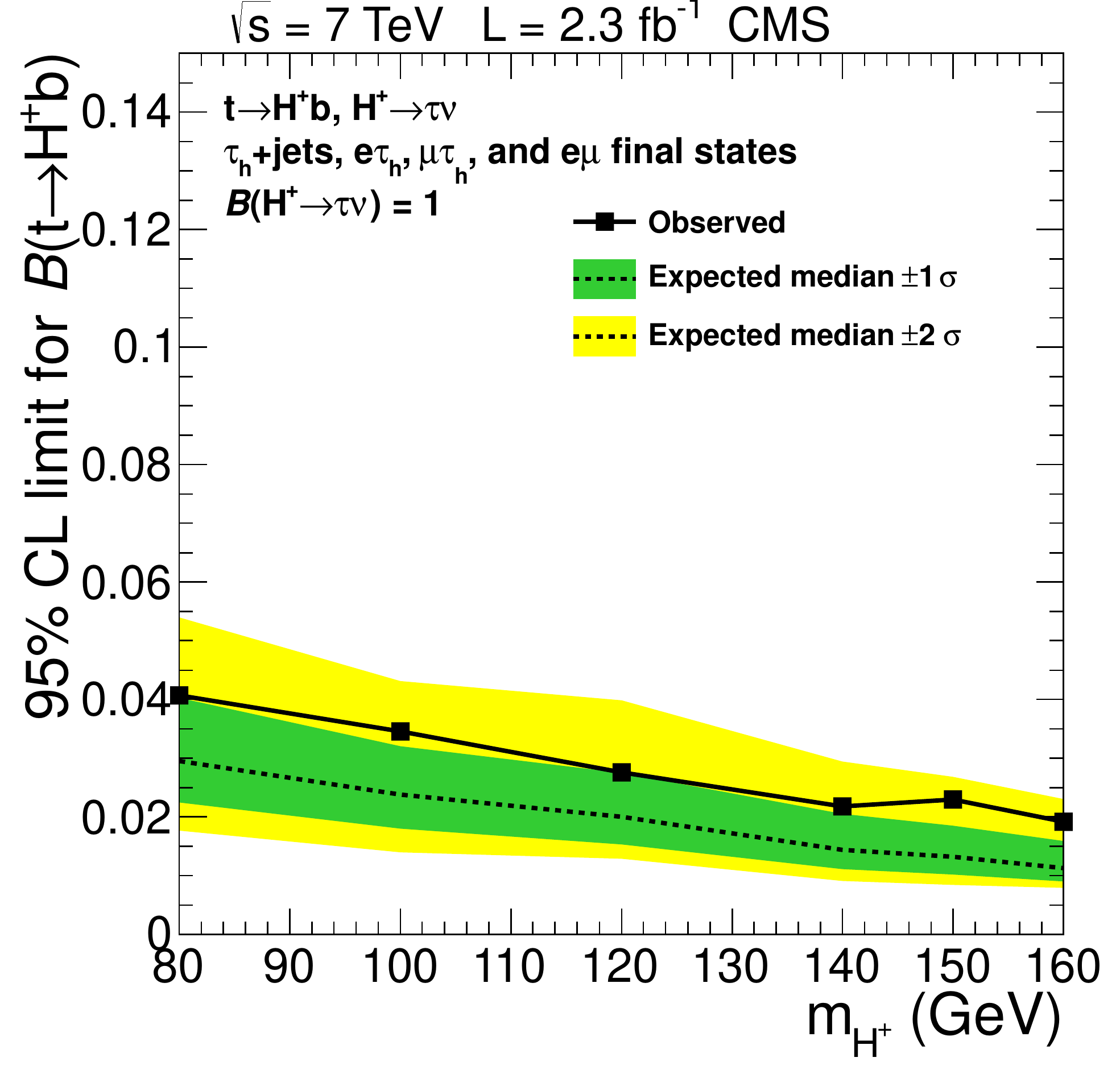}
\includegraphics[width=0.45\textwidth]{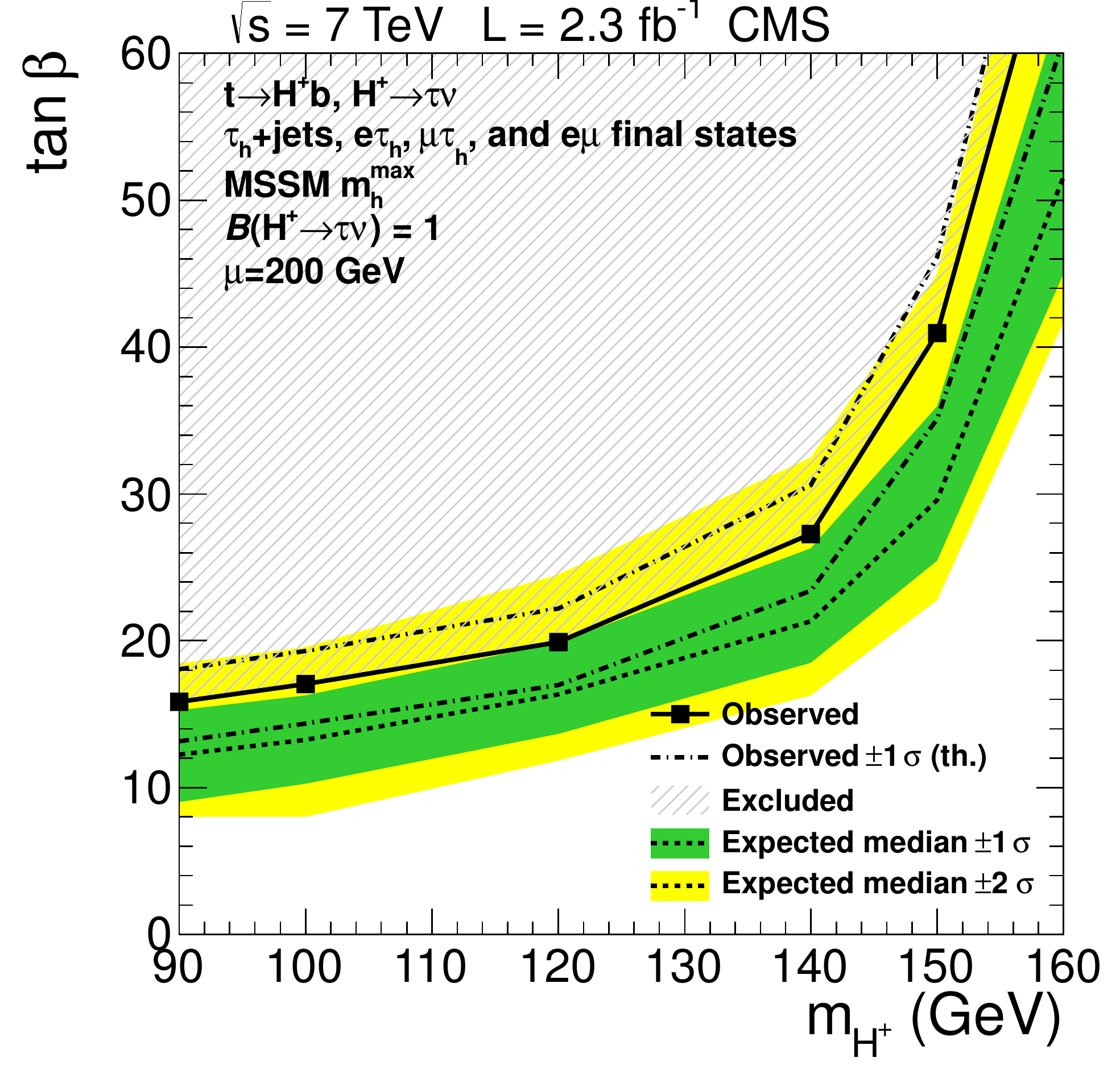}
\caption{Left: the upper limit on $\mathcal{B}(\cPqt \rightarrow \PH^{+}\cPqb)$ as a function of $m_{\PH^{+}}$
         obtained from the combination of the all final states.
         Right: the exclusion region in the MSSM $M_{\PH^{+}}$-$\tan \beta$ parameter space obtained from
         the combined analysis for the MSSM $m_\mathrm{h}^\text{max}$ scenario ~\cite{Carena:1999xa}.
         The $\pm 1 \sigma$ and $\pm 2 \sigma$ bands around the expected limit are also shown.}
\label{fig:limitcombination_fig7and8}
\end{center}
\end{figure}

The upper limit on $\mathcal{B}(\cPqt\rightarrow\PH^{+}\cPqb)$ as a function of $m_{\PH^{+}}$ is shown in Fig.~\ref{fig:limitcombination_fig3and4}
for the fully hadronic and $\Pe\Pgt_\mathrm{h}$ final states and in Fig.~\ref{fig:limitcombination_fig5and6} for the $\Pgm\Pgt_\mathrm{h}$ and $\Pe\Pgm$
final states. The combined upper limit has been obtained using the procedure described in~\cite{combination}.
Figure~\ref{fig:limitcombination_fig7and8} (left) shows the upper limit obtained from the combination of all final states.

Table~\ref{tab:limitcombination_tab1} gives the values of the median, $\pm 1 \sigma$, and $\pm 2 \sigma$ expected and the observed
95\% CL upper limit for $\mathcal{B}(\cPqt \rightarrow \PH^{+}\cPqb)$ as a function of $m_{\PH^{+}}$ for the combination of the fully hadronic,
$\Pe\Pgt_\mathrm{h}$, $\Pgm\Pgt_\mathrm{h}$, and $\Pe\Pgm$ final states.
The systematic uncertainties for the $\Pe\Pgt_\mathrm{h}$, $\Pgm\Pgt_\mathrm{h}$, and $\Pe\Pgm$ analyses are larger than the statistical uncertainties.
\begin{table*}[htbp]
\begin{center}
\topcaption{The expected range and observed 95\% CL upper limit for $\mathcal{B}(\cPqt \rightarrow \PH^{+}\cPqb)$
         as a function of $m_{\PH^{+}}$ for the combination of the fully hadronic, $\Pe\Pgt_\mathrm{h}$, $\Pgm\Pgt_\mathrm{h}$,
         and $\Pe\Pgm$ final states.}
\begin{tabular}{|c|c|c|c|c|c|c|}
  \hline
                          \multicolumn{7}{|c|}{95\% CL upper limit on $\mathcal{B}(\cPqt \rightarrow \PH^{+}\cPqb)$}    \\
\hline
\hline
  $m_{\PH^{+}}$   & \multicolumn{5}{c|}{Expected limit}                         & Observed   \\\cline{2-6}
  (\GeVns)   & $-2\sigma$  & $-1\sigma$ & median & +1$\sigma$ & +2$\sigma$  & limit       \\
  \hline
  \hline
    80          &  0.018        &  0.022       &  0.029      &  0.040       &  0.054       &  0.041               \\
  \hline
   100          &  0.014        &  0.018       &  0.024      &  0.032       &  0.043       &  0.035               \\
  \hline
   120          &  0.013        &  0.015       &  0.020      &  0.027       &  0.040       &  0.028               \\
  \hline
   140          &  0.009        &  0.011       &  0.014      &  0.021       &  0.030       &  0.022               \\
  \hline
   150          &  0.008        &  0.010       &  0.013      &  0.019       &  0.027       &  0.023               \\
  \hline
   160          &  0.008        &  0.009       &  0.011      &  0.016       &  0.023       &  0.019               \\
  \hline
\end{tabular}

\label{tab:limitcombination_tab1}
\end{center}
\end{table*}
Figure~\ref{fig:limitcombination_fig7and8} (right) shows the exclusion region in the MSSM $m_{\PH^{+}}$-$\tan \beta$
parameter space obtained from the combined analysis for the MSSM $m_\mathrm{h}^\text{max}$ scenario~\cite{Carena:1999xa}:
$M_\mathrm{SUSY} = 1\TeV$, $\mu = +200\GeV$, $M_{2} = 200\GeV$,
$m_{\PSg} = 0.8 M_\mathrm{SUSY}$, $X_{\cPqt} = 2M_\mathrm{SUSY}$, and
$A_{\cPqb} = A_{\cPqt}$.
Here, $M_\mathrm{SUSY}$ denotes the common soft-SUSY-breaking squark mass of the third generation;
$X_{\cPqt} = (A_{\cPqt} - \mu / \tan\beta)$ is the stop mixing parameter; $A_{\cPqt}$ and $A_{\cPqb}$ are the stop
and sbottom trilinear couplings, respectively; $\mu$ the Higgsino mass parameter; $M_{g}$ the gluino
mass; and $M_{2}$ is the SU(2)-gaugino mass parameter. The value of $M_{1}$ is fixed via the unification
relation $M_{1} = (5/3) M_{2} \sin \theta _{\PW} / \cos \theta _{\PW}$.

The $\cPqt \rightarrow \PH^{+}\cPqb$ branching fraction is calculated with the FeynHiggs program~\cite{Hahn:2009zz}.
The exclusion contours corresponding to the $\pm 1\sigma$ theoretical error on $\mathcal{B}(\cPqt \rightarrow \PH^{+}\cPqb)$ due to missing one-loop
EW corrections (5\%), missing two-loop QCD corrections (2\%) and  $\Delta_{b}$ induced uncertainties (the $\Delta _{b}$ term accumulates
the SUSY-QCD corrections)~\cite{Dittmaier:2012vm} are also shown in Fig.~\ref{fig:limitcombination_fig7and8} (right).

The upper limit on the the branching fraction $\mathcal{B}(\cPqt\rightarrow \PH^{+}\cPqb)$ and the exclusion region in the
MSSM $m_{\PH^{+}}$-$\tan \beta$ parameter space obtained from the combined analysis are
comparable with the results from the ATLAS experiment~\cite{atlasH+}.
\section{Summary \label{sec:conclusion}}

   A search has been performed for a light charged Higgs boson produced in top quark decays $\cPqt\rightarrow \PH^{+}\cPqb$
   and which in turn decays into $\Pgt^{+} \Pgngt$. The data sample used in the analysis corresponds to an integrated luminosity of
   about 2\fbinv. The fully hadronic, $\Pe\Pgt_\mathrm{h}$, $\Pgm\Pgt_\mathrm{h}$, and $\Pe\Pgm$ final states have been used in the analysis.
   The results from these analyses have been combined to extract limits on $\cPqt\rightarrow \PH^{+}\cPqb$
   branching fraction. Upper limits on the branching fraction $\mathcal{B}({\cPqt} \rightarrow \PH^{+} {\cPqb})$ in the range of 2--4\% are
   established for charged Higgs boson masses between 80 and 160\GeV, under the assumption that
   $\mathcal{B}(\PH^{+} \rightarrow \Pgt^{+} \Pgngt) = 1$.

\section*{Acknowledgements \label{sec:acknowledgements}}
We congratulate our colleagues in the CERN accelerator departments for the excellent performance of the LHC machine. We thank the technical and administrative staff at CERN and other CMS institutes, and acknowledge support from: FMSR (Austria); FNRS and FWO (Belgium); CNPq, CAPES, FAPERJ, and FAPESP (Brazil); MES (Bulgaria); CERN; CAS, MoST, and NSFC (China); COLCIENCIAS (Colombia); MSES (Croatia); RPF (Cyprus); MoER, SF0690030s09 and ERDF (Estonia); Academy of Finland, MEC, and HIP (Finland); CEA and CNRS/IN2P3 (France); BMBF, DFG, and HGF (Germany); GSRT (Greece); OTKA and NKTH (Hungary); DAE and DST (India); IPM (Iran); SFI (Ireland); INFN (Italy); NRF and WCU (Korea); LAS (Lithuania); CINVESTAV, CONACYT, SEP, and UASLP-FAI (Mexico); MSI (New Zealand); PAEC (Pakistan); MSHE and NSC (Poland); FCT (Portugal); JINR (Armenia, Belarus, Georgia, Ukraine, Uzbekistan); MON, RosAtom, RAS and RFBR (Russia); MSTD (Serbia); MICINN and CPAN (Spain); Swiss Funding Agencies (Switzerland); NSC (Taipei); TUBITAK and TAEK (Turkey); STFC (United Kingdom); DOE and NSF (USA).

Individuals have received support from the Marie-Curie programme and the European Research Council (European Union); the Leventis Foundation; the A. P. Sloan Foundation; the Alexander von Humboldt Foundation; the Belgian Federal Science Policy Office; the Fonds pour la Formation \`a la Recherche dans l'Industrie et dans l'Agriculture (FRIA-Belgium); the Agentschap voor Innovatie door Wetenschap en Technologie (IWT-Belgium); the Council of Science and Industrial Research, India; and the HOMING PLUS programme of Foundation for Polish Science, co-financed from European Union, Regional Development Fund.

\bibliography{auto_generated}   

\newpage
\cleardoublepage \appendix\section{The CMS Collaboration \label{app:collab}}\begin{sloppypar}\hyphenpenalty=5000\widowpenalty=500\clubpenalty=5000\textbf{Yerevan Physics Institute,  Yerevan,  Armenia}\\*[0pt]
S.~Chatrchyan, V.~Khachatryan, A.M.~Sirunyan, A.~Tumasyan
\vskip\cmsinstskip
\textbf{Institut f\"{u}r Hochenergiephysik der OeAW,  Wien,  Austria}\\*[0pt]
W.~Adam, E.~Aguilo, T.~Bergauer, M.~Dragicevic, J.~Er\"{o}, C.~Fabjan\cmsAuthorMark{1}, M.~Friedl, R.~Fr\"{u}hwirth\cmsAuthorMark{1}, V.M.~Ghete, J.~Hammer, N.~H\"{o}rmann, J.~Hrubec, M.~Jeitler\cmsAuthorMark{1}, W.~Kiesenhofer, V.~Kn\"{u}nz, M.~Krammer\cmsAuthorMark{1}, D.~Liko, I.~Mikulec, M.~Pernicka$^{\textrm{\dag}}$, B.~Rahbaran, C.~Rohringer, H.~Rohringer, R.~Sch\"{o}fbeck, J.~Strauss, A.~Taurok, W.~Waltenberger, G.~Walzel, E.~Widl, C.-E.~Wulz\cmsAuthorMark{1}
% ...
\vskip\cmsinstskip
\dag:~Deceased\\
1:~~Also at Vienna University of Technology, Vienna, Austria\\
% ...

\end{sloppypar}
\end{document}